\documentclass[12pt]{article}
\usepackage[utf8]{inputenc}
\usepackage[margin=0.75in]{geometry}
\usepackage[numbers,sort&compress]{natbib}
\usepackage[usenames,dvipsnames]{xcolor}
\definecolor{Salmon}{RGB}{250,128,114}
\usepackage[colorlinks=true, linkcolor=Salmon, citecolor=Salmon, urlcolor=Salmon]{hyperref} %
\usepackage{bbold}
\usepackage{sidecap}
\usepackage{amsmath}
\usepackage[affil-it]{authblk}
\usepackage{booktabs}
\usepackage{setspace}
\usepackage{todonotes}
\usepackage{comment}
\usepackage{amsfonts}
\usepackage{amssymb}
\usepackage{amsthm} 
\usepackage{multicol}
\usepackage{graphicx}
\usepackage{enumerate}
\usepackage[inline,shortlabels]{enumitem}
\usepackage{bbm}
\usepackage{bm}
\usepackage{fancyhdr}
\usepackage{floatpag}
\usepackage{setspace}

\def\vec#1{{\bm #1}}

\def\vec#1{{\bm #1}}

\def\mat#1{\mathbf{#1}}

\def\given{\mid}

\definecolor{darkred}{rgb}{0.55, 0.0, 0.0}

\title{Uncovering the universal dynamics of citation systems: From science of science to law of law and patterns of patents}
\title{Modeling Long-Term Citation Patterns across Science, Law, and Patents}
\title{Modeling Collective/Community Citation Patterns across Science, Law, and Patents}
\title{Community-centric modeling of citation dynamics explains collective citation patterns in science, law, and patents}

\author[1,7,*]{Sadamori Kojaku}
\author[2,3,4,*]{Robert Mahari}
\author[4,5,+]{Sandro Claudio Lera}
\author[4,6]{Esteban Moro}
\author[2,4]{Alex Pentland}
\author[4,7,+]{Yong-Yeol Ahn}

\affil[1]{Department of Systems Science and Industrial Engineering, Binghamton University, New York, USA}
\affil[2]{Media Lab, Massachusetts Institute of Technology, Cambridge, USA}
\affil[3]{Harvard Law School, Cambridge, USA}
\affil[4]{Connection Science, Massachusetts Institute of Technology, Cambridge, USA}
\affil[5]{Institute of Risk Analysis, Prediction and Management (Risks-X), Southern University of Science and Technology, Shenzhen, China}
\affil[6]{Network Science Institute, Department of Physics, Northeastern University, Boston, USA}
\affil[7]{Center for Complex Networks and Systems Research, Luddy School of Informatics, Computing, and Engineering, Indiana University, Bloomington, USA}
\affil[*]{Authors contributed equally}
\affil[+]{Corresponding authors: slera@mit.edu and yyahn@iu.edu}

\date{\today}

\begin{document}

\maketitle
\begin{abstract}
Many human knowledge systems, such as science, law, and invention, are built on documents and the citations that link them. Citations, while serving multiple purposes~\cite{small1978cited, latour2013laboratory, Price1976}, primarily function as a way to explicitly document the use of prior work and thus have become central to the study of knowledge systems~\cite{garfield1955citation, price1965networks, shapiro1992origins}.
Analyzing citation dynamics has revealed statistical patterns that shed light on knowledge production, recognition, and formalization~\cite{price1963little, price1965networks, barabasi1999, candia2019universal,candia2020rising, radicchi2008universality, milojevic2015quantifying, Ke2015defining, fortunato2018science}, and has helped identify key mechanisms driving these patterns~\cite{wang2013quantifying, price1965networks, Sinatra2016-oi}. 
However, most quantitative findings are confined to scientific citations, raising the question of universality of these findings.
Moreover, existing models of individual citation trajectories fail to explain phenomena such as delayed recognition, calling for a unifying framework. 
Here, we analyze a newly available corpus of U.S. case law, in addition to scientific and patent citation networks, to show that they share remarkably similar citation patterns, including a heavy-tailed distribution of sleeping beauties. We propose a holistic model that captures the three core mechanisms driving collective dynamics and replicates the elusive phenomenon of delayed recognition. 
We demonstrate that the model not only replicates observed citation patterns, but also better predicts future successes by considering the whole system. Our work offers insights into key mechanisms that govern large-scale patterns of collective human knowledge systems and may provide generalizable perspectives on discovery and innovation across domains.
\end{abstract}

\section{Introduction}
\label{sec:introduction}

Human culture is built on the ability to record and accumulate knowledge and experience as well as the capacity to use those accumulated experiences to negotiate and resolve conflicts.
These abilities have become formalized into \emph{knowledge systems} such as science and law. 
Human knowledge systems are often built upon \emph{publications} and \emph{citations} where each publication represents a unit of facts, knowledge, or argument, and citations formally document the use of prior works.


Since citations are fundamental to the evolution of knowledge systems, they have been indexed, counted, and analyzed~\cite{garfield1955citation, garfield1957breaking, small1973co, small1985clustering, kessler1963bibliographic, fortunato2018science}. 
Examining how scientists publish and cite prior work has revealed both nomothetic and idiographic insights into collective formalization of scientific knowledge.
Citations have been found to serve a number of functions, including as the result of practical incentives to grow research labs~\cite{latour2013laboratory}, as the basis for arguments that resolve disagreements~\cite{teufel1999argumentative}, and as a currency that underpins scientific reputation and status~\cite{merton1973sociology}.
More quantitative research seeks to measure statistical properties of citation systems, which also facilitates quantitative predictions about the future trajectory of publications and fields.
For instance, it has been shown that the number of new citations tends to be proportional to a publication's existing number of citations (known as preferential attachment, the Matthew Effect, or ``the rich get richer'')~\cite{price1965networks, barabasi1999}.
At the same time, new knowledge attracts attention while older knowledge is gradually forgotten, which leads to the recency bias, a tendency to cite recent publications over old counterparts, contributing to the integration of novel discoveries into our knowledge base~\cite{candia2019universal, wang2013quantifying, candia2020rising}.
Meanwhile, some publications (``sleeping beauties'') go unnoticed for decades before experiencing a surge in attention~\cite{van2004sleeping, stent1972prematurity, zuckerman1980indicators, Ke2015defining}.
Although there have been indications that citation patterns may be universal across citation-based knowledge systems, system-level data and measurements, particularly for legal systems, have not been readily available. 
Furthermore, the delayed recognition phenomenon has been elusive to model within the current paradigm of citation models that are built on preferential attachment and collective forgetting because these directly exclude the possibility of discovering ``sleeping beauties''.


Here, we leverage a newly available dataset of U.S. federal judicial opinions, and the associated citation network, to demonstrate that three knowledge systems fundamental to human civilization---science, law, and invention---exhibit remarkably similar large-scale patterns. We propose a holistic model that succinctly captures the key mechanisms that are sufficient to produce these shared patterns, resolving the paradox of sleeping beauties along the way.

Scientists produce new knowledge by building on prior contributions~\cite{garfield1955citation, price1965networks, shapiro1992origins, merton1968social}.
In `common law' jurisdictions, which cover a third of the global population~\cite{fathally2009systemes}, judges and lawyers base their legal reasoning on citations to prior judicial decisions (precedent).
As a result, the common law constantly evolves as judges issue new decisions (see SI Appendix~\ref{SI:us_common_law} for an overview of the US federal law system).
For over 150 years, U.S. jurists have used citations to systematize the precedent landscape but only relatively recently have these citations been analyzed in a quantitative manner at a large scale~\cite{shapiro1992origins}.
In patent applications, inventors cite prior art to contextualize their inventions and to distinguish their contributions from past inventions.
Patent examiners use these citations to determine whether an invention reaches the threshold for patentability and may add their own citations to prior art.

While science, patents, and law are collective knowledge systems built on citations, they are distinct in terms of their incentives, norms, and procedures (see SI Appendix~\ref{SI:compare_knowledge_systems}).
For instance, anyone can, in principle, publish in scientific journals or file a patent, whereas one has to be elected or appointed as a judge to publish a judicial opinion.
Scientists and inventors usually \emph{choose} what to work on while judges are \emph{assigned} to cases at random.
The number of scientists and inventors has been growing rapidly while the number of U.S. federal judges has remained relatively stable.
Inventions are primarily driven by commercial interests, unlike science and law.
The U.S. legal system features a codified, binding hierarchy (e.g., Supreme Court decisions supersede all other precedents), while other systems do not.
These differences prompt an examination of the generalizability of the ``laws of citation'' observed in science and an exploration of whether the evolution and formalization of knowledge systems through citations follow similar patterns across different systems.

Here, we show that, despite the differences between how these systems produce knowledge, the fundamental statistical citation dynamics are remarkably similar in terms of characteristics, including preferential attachment, exponential growth, citation recency, and delayed recognition (sleeping beauties).
Based on this, we argue that the formalization of knowledge may be largely shaped by intrinsic human constraints and robust basic mechanisms, and the resulting statistical patterns are robust against numerous factors that distinguish individual knowledge systems.
Although current models of citation dynamics can capture some of these patterns~\cite{wang2013quantifying,price1965networks, barabasi1999}, they are unable to account for all of them.
While these existing models focus on \emph{individual publications that receive citations}, we argue that capturing the dynamics of the \emph{entire knowledge communities that cite} is the key to understanding the dynamics of knowledge systems as a whole.
Thus, we introduce the \emph{Community Citation Model} (CCM) that describes how an entire knowledge system collectively evolves, identifies, and builds on existing knowledge.
The CCM sheds light on the quantitative mechanisms behind the creation of new knowledge and how important contributions can be discovered after being unnoticed for a long time.
We show that our model not only explains the aforementioned patterns but also better predicts the future citation success of publications by holistically modeling the evolution of the whole system.

\section{Results}
\label{sec:results}

\begin{figure}
    \centering
    \includegraphics[width=0.75\textwidth]{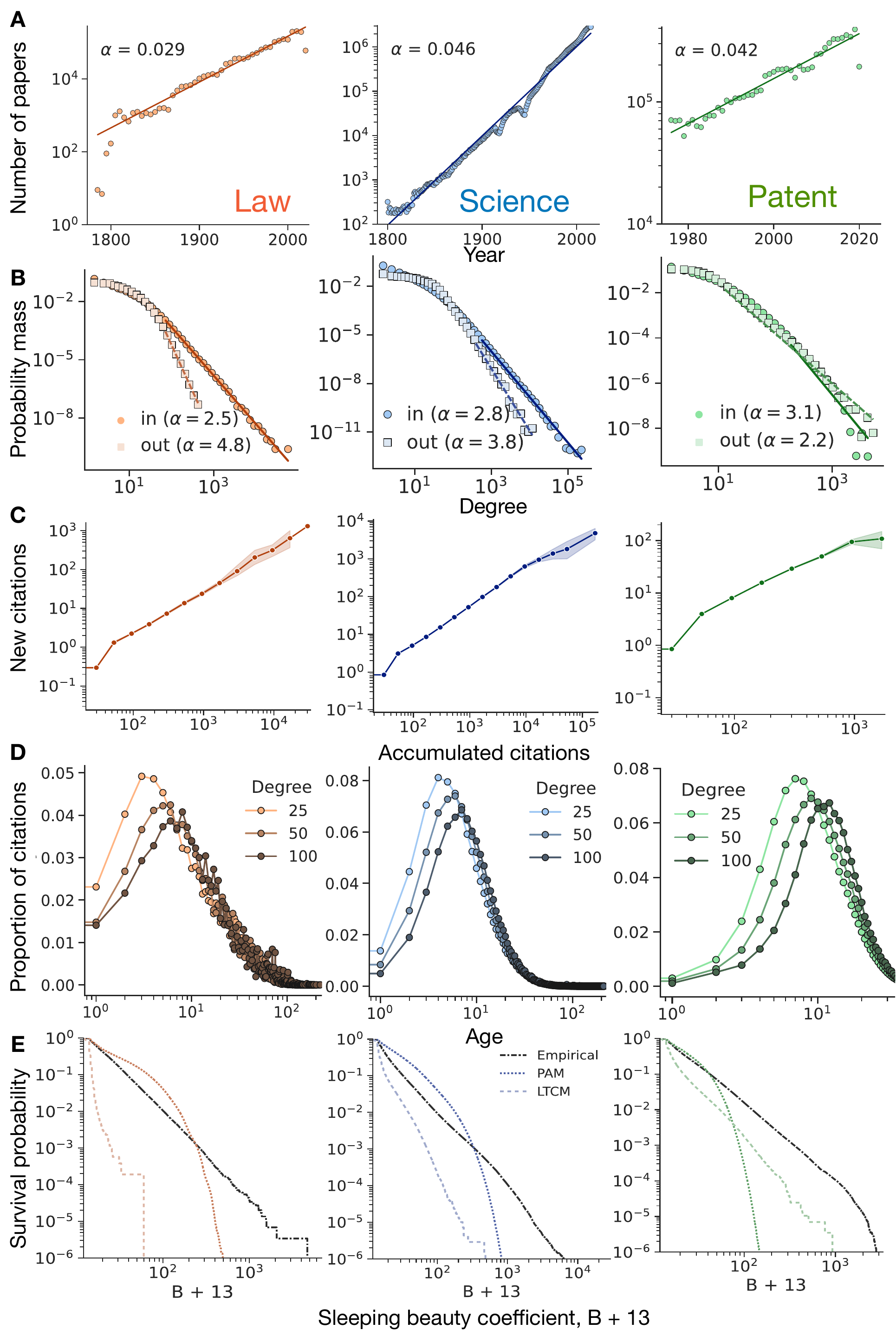}
    \caption{
        Universal patterns of citation dynamics across three knowledge systems (law, science, and patents).
        {\bf A}. Exponential growth in publications.
        {\bf B}. Degree distribution of the publications. 
        {\bf C}. Linear preferential attachment, where the number of new citations acquired in a year is proportional to the accumulated number of citations.
        {\bf D}. Aging function representing the probability that a publication receives the $x+1$th citation at age $\Delta t$, where $x \in \{25, 50, 100\}$.
        {\bf E}. Skewed distribution of the Sleeping Beauty coefficient ($B$).
        }
    \label{fig:citation-dynamics}
\end{figure}

\subsection{Universal citation patterns}
\label{sec:universal_patterns}

To investigate the universality of statistical citation patterns, we examine three distinct knowledge systems: science, patents, and common law.
For scientific citations, we primarily use the comprehensive SciSciNet dataset, though we also employ the American Physical Society (APS) dataset when conducting computationally intensive analyses due to the large size of SciSciNet dataset (results from SciSciNet are labeled as ``Science,'' while those from the APS dataset are labeled as ``APS'').
For patent citations, we examine data from the United States Patent and Trademark Office (USPTO).
These citation datasets are extensively studied and well documented~\cite{price1965networks, barabasi1999, candia2019universal, candia2020rising, Ke2015defining, wang2013quantifying,park2023papers,funk2017dynamic,marx2020reliance} 
(see Section~\ref{sec:data_sources} for data preprocessing).
In the legal domain, we concentrate on the U.S. federal system as a good representation of judicial citation dynamics, including reliance on precedent and the binding hierarchy of legal decisions (SI Appendix~\ref{SI:us_common_law}).
While citation patterns in science and patents have been subject to extensive study, less work has compared patterns across different systems or focused on statistical dynamics of legal citations.
Examining fundamental citation patterns that are well-known from science, including citation growth, preferential attachment, recency, and delayed recognition, we find that all three systems demonstrate strikingly similar dynamics (see Fig.~\ref{fig:citation-dynamics}) despite their different \emph{modi operandi} (see SI Appendix~\ref{SI:compare_knowledge_systems}).

First, all three systems display exponential growth in the number of publications and citations (see Fig.~\ref{fig:citation-dynamics}A and Fig.~\ref{fig:growth}A).
Although this phenomenon has been documented in the realm of science since at least 1965~\cite{price1965networks}, it is noteworthy to observe this growth in the legal domain, considering that the number of U.S. federal judges has remained relatively constant over time.
Second, the number of references cited by each publication also exhibits exponential growth across all systems, reflecting the growing amount of relevant knowledge that is required to produce new knowledge or perhaps the growing ``burden of knowledge''~\cite{jones2009burden} (see Fig.~\ref{fig:growth}B).
Our observation is inconsistent with a previous study~\cite{krapivsky2005network} that claimed the number of references in science grows logarithmically rather than exponentially. 
Our validation showed that the logarithmic growth fits poorly, especially for law and patents, compared to the exponential growth (SI Appendix~\ref{SI:ref_growth}). 
Third, citation networks in all three systems display power-law degree distributions, and citation dynamics exhibit preferential attachment~\cite{Price1976, barabasi1999} (Fig.~\ref{fig:citation-dynamics}B and C).
Fourth, even though publications tend to accumulate citations over time due to preferential attachment, older publications become less likely to be cited as their novelty and relevance diminish~\cite{candia2019universal, Medo2011-ko, wang2013quantifying}.
In fact, the probability that a publication receives the $x+1$-th citation at age $\Delta t$ peaks within the first 10 years and decreases over time (Fig.~\ref{fig:citation-dynamics}D; see Section~\ref{sec:exponential_growth} for how to compute this probability). 
This pattern of obsolescence can be found across all systems.
Finally, we observe that all systems exhibit unexpectedly abundant delayed recognition of some publications, a phenomenon commonly referred to as ``Sleeping Beauties'' or ``delayed recognition''~\cite{zuckerman1980indicators, Ke2015}.
As shown in Fig.~\ref{fig:citation-dynamics}E, the ``Sleeping Beauty coefficient'' $B$~\cite{Ke2015defining}, which captures the intensity of a publication's delayed recognition, has heavier tails than what would be expected based on preferential attachment alone.
A striking example of this phenomenon is the 1935 article by Albert Einstein, Boris Podolsky, and Nathan Rosen, which garnered limited attention for nearly 70 years before witnessing a citation explosion.
Similarly, the Court of Appeals case \textit{United States v. Hooton}, in which an appeal was dismissed for being insubstantial, was virtually uncited for two decades before being discovered (or ``awakened'') and cited explosively.

Our findings reveal that all three systems display an extensive set of universal macroscopic citation dynamics.
The observed universality of citation patterns implies that, despite the numerous distinct characteristics of knowledge generation across systems, the process of knowledge formalization through citations may be predominantly universal.

\subsection{Modeling universal citation dynamics}
\label{sec:baseline}

\begin{figure}[h!p]
        \centering
        \includegraphics[height=\dimexpr \textheight - 7\baselineskip\relax]{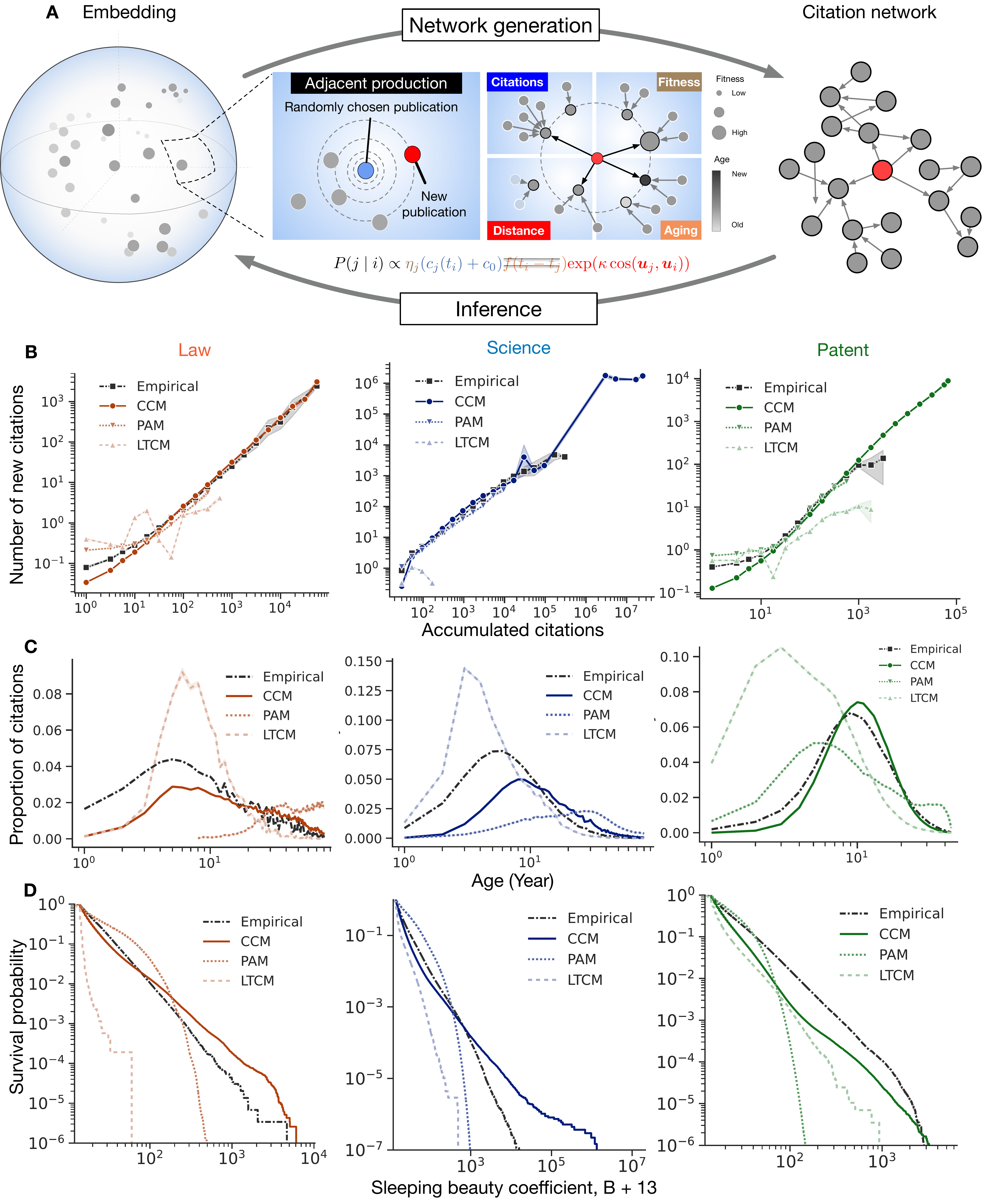}
        \caption{
        The Community Citation Model (CCM) reproduces universal citation patterns.
        {\bf A}. A publication is represented as a point in a vector space.
        A new publication is placed near a randomly chosen existing publication and cites nearby publications according to Eqs.~\eqref{eq:adj_prod} and \eqref{eq:CCM}.
        {\bf B--D}. The CCM reproduces preferential attachment, recency, and the skewed sleeping beauty coefficient distribution. In 
        {\bf C} we show the aging function for publications with 50 references.
        We compare the CCM to the preferential attachment model (PAM) and the long-term citation model (LTCM).
        }
        \label{fig:reproduced-citation-characteristcs}
\end{figure}

A prevailing explanation of universal citation patterns is captured by the long-term citation model (LTCM)~\cite{wang2013quantifying, shenModelingPredictingPopularity2014}.
The LTCM posits that the number of citations a publication receives depends on the following three features of the publication itself: the publication's intrinsic quality (`fitness'), its recency (`aging'), and the number of accumulated citations (`preferential attachment').
Together, these components describe the probability that publication $i$ cites publication $j$:
\begin{equation}
    P_{\text{LTCM}}(j \given i) = \frac{1}{Z(t_i)}~ (c_j(t_i) + c_0)~  \eta_j ~ f( t_i - t_j\; ;\mu_j, \sigma_j),
    \label{eq:LTCM}
\end{equation}
where $c_j(t_i)$ represents the number of citations that publication $j$ has received by the time $t_i$ when publication $i$ is published, and $c_0$ is a pre-defined parameter to offset citations to prevent the citation probability from being zero. We set $c_0 = 30$ per the original paper~\cite{wang2013quantifying}.
Fitness $\eta_j >0$ signifies the publication's intrinsic quality.
The denominator, $Z(t_i)$, is a normalization constant.
The function $f( \cdot )$ is a log-normal distribution characterizing the aging effect,
\begin{equation}
    f(t - t_j\;;\mu_j, \sigma_j) = \frac{1}{\sigma_j\sqrt{2\pi (t-t_j)}}\exp\left[-\frac{\left(\log (t - t_j) - \mu_j\right)^2}{2\sigma^2 _j}\right]\;\quad (t > t_j),
\end{equation}
parameterized by \emph{publication-level} mean $\mu_j$ and variance $\sigma_j$ parameters specific to publication $j$.
Naturally, the probability of citing a publication before it is published is zero, i.e., $f(t - t_j)=0$ if $t \leqslant t_j$.
The LTCM is reduced to the preferential attachment model (PAM) if the recency $f( \cdot )$ and intrinsic quality $\eta_j$ are constant across all publications and time:
\begin{equation}
    P_{\text{PAM}}(j \given i) \propto (c_j(t_i) + c_0).
    \label{eq:PAM}
\end{equation}

Although the LTCM has been evaluated with scientific citations, it is inherently difficult for the model to explain certain phenomena like sleeping beauties that deviate from the assumed aging function~\cite{wang2013quantifying,shenModelingPredictingPopularity2014}.
To explore how well it explains the patterns observed across all three knowledge systems, we fit the LTCM to science, legal, and patent citation networks using maximum likelihood estimation~\cite{shenModelingPredictingPopularity2014} and regenerate the citation trajectories of publications (See Section~\ref{sec:long_term_citation_model}).
Similarly, we fit the PAM to the three networks.
The LTCM reproduces preferential attachment and recency patterns for all three systems (Fig.~\ref{fig:reproduced-citation-characteristcs}).
However, by construction, neither the LTCM nor PAM give rise to enough sleeping beauties because the probability of ``discovering'' a relatively uncited publication after a long time is low in both models.
The high fitness necessary for the ``awakening'' is at odds with a publication experiencing delayed recognition.
This implies that neither model accounts for the discovery process that appears to be at work across knowledge systems.

Examining the models' implicit assumptions reveals that the LTCM and PAM share the same perspective on success (in terms of citations): in both models, the probability of citation is entirely determined by the \emph{attributes of the publication itself}.
In other words, these models completely overlook the collective context or knowledge landscape of existing publications.
Of course, this is far from reality, where both the production of knowledge and citation decisions are heavily influenced by the community's collective dynamics.
These collective dynamics influence the areas in which new knowledge is produced and which existing publications are discovered.
As a major departure from the existing models, we consider that citations to a publication are not solely a result of the \textit{cited} publication's attributes, but rather a reflection of collective decisions by the authors of the \textit{citing} publications.
Our idea echoes the adage: ``Success isn't about \emph{you}, it's about \emph{us}''~\cite{barabasi2018formula}.

Our empirical results described in Section \ref{sec:universal_patterns} already hint at these collective phenomena.
Sleeping beauties, for instance, are fundamentally impossible to be explained by models that only take the cited publication's attributes into account.
By contrast, they can be naturally explained by a knowledge community that shifts its attention, explores yet undiscovered knowledge, and finally ``discovers'' neglected publications.

Our new model of knowledge formalization---the \emph{Community Citation Model} (CCM)---builds on this idea that collective attention, and the dynamics of collectives are the key to capturing the dynamics of knowledge systems.
To this end, we imagine a \emph{space of knowledge} containing all possible knowledge and corresponding publications.
Intuitively, new publications are most likely to emerge in ``hot'' areas---dense and fast-growing regions of knowledge space, also known as ``emerging research fronts''~\cite{upham2010emerging}.
Occasionally, however, a prescient publication may appear in an empty space that has yet to be explored.
If this publication was ``fit'' but ``too early'',  it can go unnoticed for a long time, until the whole community finally moves to the region and discovers the publication.
This scenario highlights how citations depend not only on publications' inherent characteristics but also on their relationship to \emph{where} the rest of the community is located within the knowledge space.

Operationally, we represent the space of knowledge by a $K$-dimensional spherical space (Fig.~\ref{fig:reproduced-citation-characteristcs}A).
Unit vector $\vec{u}_i$ represents the location of publication $i$ in the knowledge space.
We propose a zero-knowledge baseline model that randomly places a new publication by first picking an existing publication $j$ uniformly at random and then placing a new publication $i$ around $j$ based on the von Mises distribution centered at $j$, i.e.,
\begin{align} \label{eq:adj_prod}
        P(\vec{u_j} \given  \vec{u_i}) \propto \exp\left( -\lambda d( \vec{u_i}, \vec{u_j} ) \right),
\end{align}
where $\lambda$ is a scaling parameter that controls the closeness between the new publication $i$ and the existing publication $j$, in terms of the cosine distance, $d( \vec{u_i}, \vec{u_j} ) = 1- \vec{u_i} \cdot \vec{u_j}$.
When $\lambda$ is large, new publications tend to appear in closer proximity to existing publications, leading to strongly concentrated clusters.
As $\lambda$ decreases, the concentration becomes less pronounced, and publications are more spread out.
As a new publication is placed, it cites existing publication $j$ according to the following probability model:
\begin{align}
        P_{\text{CCM}}( j \given i; \vec{u}_i, \vec{u}_j)  =\frac{1}{Z(t_i)}~ (c_j(t_i) + c_0)~  \eta_j ~ \exp(-\kappa d(\vec{u}_i, \vec{u}_j)),
         \label{eq:CCM}
\end{align}
where $\kappa$ is a scaling parameter that controls the contribution of the closeness to the citation probability (see Methods for details).
We set $c_0 = 20$ and verify that different values of $c_0$ yield similar results (see SI Appendix~\ref{sec:sensitivity_analysis_reproduceability})

The fundamental shift that distinguishes our model from the LTCM is that now the citation probability is defined from the perspective of the \emph{citing} publication. While the LTCM's probability model is solely dependent on the properties of the cited publication, our model considers the relevance of existing citations based on the location of the \emph{citing} publications. 
Another crucial distinction is the lack of an explicit aging mechanism in the model. 
To capture the aging dynamics (recency bias), the LTCM must prescribe an explicit aging function and estimate aging parameters for each publication.
We find that, however, the aging factor $f$ is not necessary to reproduce the recency bias (Fig.~\ref{fig:reproduced-citation-characteristcs}C), and incorporating the aging factor into the CCM does not significantly alter the results, except for a substantial reduction of sleeping beauties (see SI Appendix~\ref{sec:sensitivity_analysis_aging}).
This suggests that a publication's aging may not be an inherent mechanism but is confounded by other, more fundamental mechanisms. 
The CCM's ability to reproduce the recency bias without explicit aging can be attributed to collective attention;
new publications are more likely to be produced in a dense area of the knowledge space.
As the focus of research communities drifts in the knowledge space, the citation probability for older publications decreases. 
This mechanism, together with the exponential growth of publications, gives rise to the recency bias.
The CCM thus offers a powerful explanation of how aging of knowledge products can occur without explicitly implementing it into the model, further underscoring the importance of modeling collective citation dynamics.


Modeling the production of new publications and the creation of citations allows us to \emph{simulate} the collective evolution of an entire knowledge system, rather than the citation trajectories of individual publications.
We fit the CCM using maximum likelihood estimation (Section~\ref{sec:CCM_fitting}) to generate citation networks.   
The generated citation networks closely mirror the empirical networks. 
They not only reproduce the preferential attachment and recency patterns observed across all three systems (Figs.~\ref{fig:reproduced-citation-characteristcs}B and C), but also reproduce the relative abundance of ``sleeping beauties'' (Fig.~\ref{fig:reproduced-citation-characteristcs}D).

\begin{figure}[ht!]
    \centering
        \includegraphics[width=0.85\textwidth]{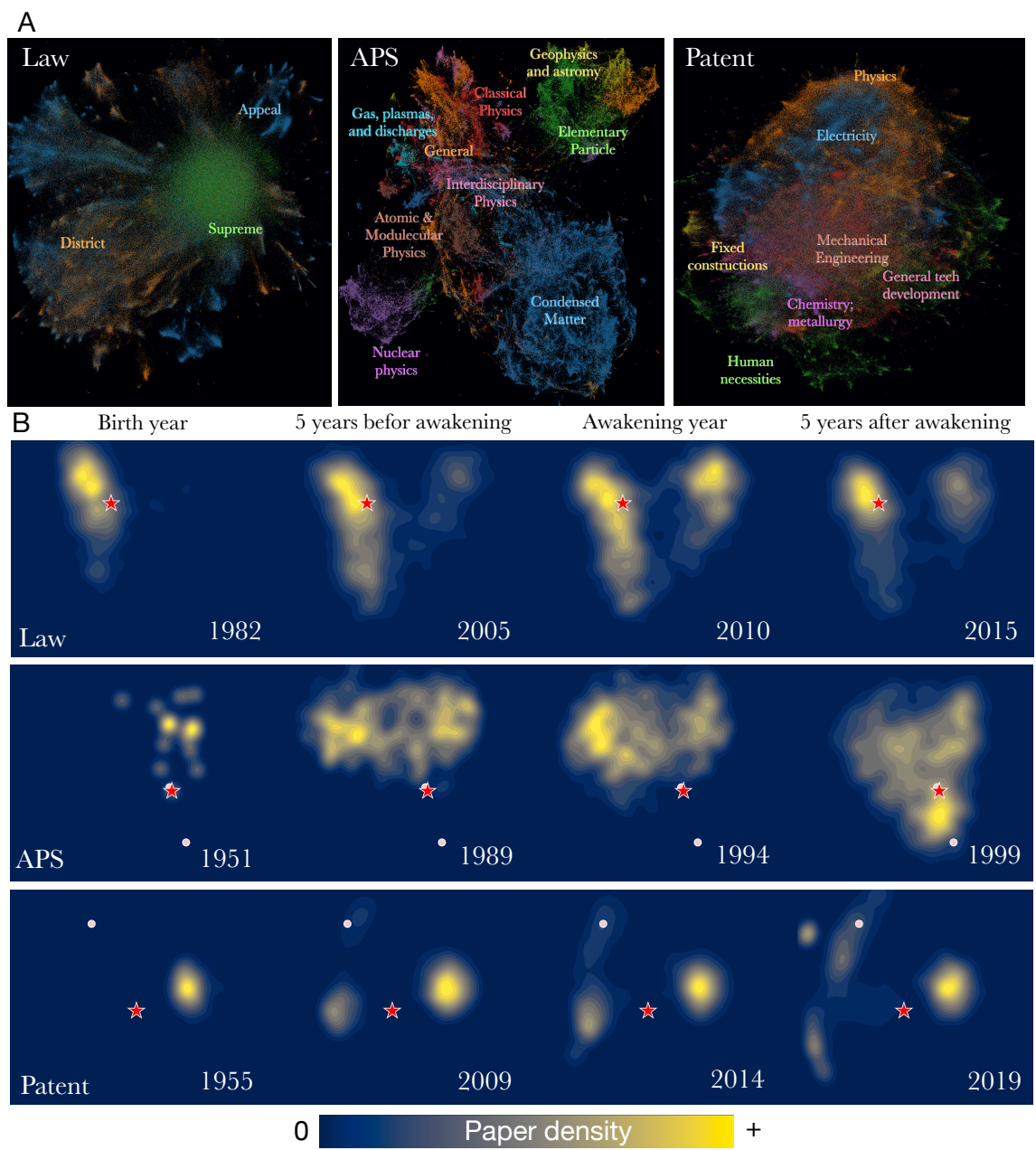}
        \caption{
        Embeddings learned by the CCM reveal community structures that help explain universal citation dynamics.
        {\bf A}. 2D projection of the 128-dimensional embedding by UMAP.
        The projection is generated with ten nearest neighbors and cosine similarity as the proximity metric.
        {\bf B}. The evolution of the embedding space in the neighborhood of the top sleeping beauties (SB).
        The brighter yellow areas represent dense areas of knowledge space in the five years leading up to the date displayed, which we approximate via kernel density estimation.
        We define the neighborhood as the 3,000 publications that are closest to the focal SBs marked by stars.
        The red circles represent other SBs in the vicinity.
        The embeddings are projected into 2D using Principal Component Analysis.
        }
        \label{fig:embedding}
\end{figure}

How and why can the CCM produce more sleeping beauties? 
An examination of some of the most pronounced ``sleeping'' publications provides insight into the underlying mechanism that is captured by the CCM (Fig.~\ref{fig:embedding}B).
Many sleeping beauty publications tend to appear in sparsely occupied niches in the knowledge space, far away from the mainstream, and as a result, they are unlikely to be cited.
Over time, however, knowledge communities evolve and drift in the knowledge space, stumbling upon a sleeping publication.
As a result, an initially neglected publication may, over time, be placed at the vanguard of a growing field.
We find evidence of this ``ahead of its time'' mechanism across all three systems (Fig.~\ref{fig:embedding}B).
This explanation highlights why modeling the collective dynamics rather than individual publication is the key to capturing this phenomenon.

A by-product of the CCM is the knowledge embedding space, where we can examine how published knowledge is organized in the high-dimensional space and build on prior work that seeks to map scientific knowledge~\cite{boyack2005mapping, borner2003visualizing}.
Despite the observed universality of empirical citation patterns, we find that the three domains of knowledge feature different types of community structure (Fig~\ref{fig:embedding}A).
In law, we find publications to be organized primarily into three communities, aligned with the hierarchy of U.S. federal courts.
By contrast, in APS and patents, the structure seems to be driven by topics, with a stronger separation in science.

The CCM also provides insights into why the number of references in publications might have been growing across all three systems (see Fig.~\ref{fig:growth}B).
Based on the CCM, we expect that references will typically be within some relatively small radius of the citing publication's position in the knowledge space.
If the radius of citations decreases, reference growth can be understood as the result of a form of specialization, where publications are citing more of their close neighbors.
We find that, in APS and patents, the radius of citation---defined by the average distance between a publication and its references---has decreased while the number of references has grown (see SI Appendix~\ref{SI:ref_growth}).
This suggests that specialization and ``the burden of knowledge'' may be increasing~\cite{jones2009burden}.
However, the citation radii appear more stable in law, which may result from the fact that U.S. judges, unlike scientists and inventors, are generalists.

Overall, by focusing on the collectives that \emph{cite} publications rather than focusing on the intrinsic attributes of \emph{cited} publications, the CCM elucidates a plausible mechanism underlying sleeping beauties and patterns related to increased specialization, which cannot be understood by publication-centric models.

\subsection{Forecasting citation trajectories}
\label{sec:forecasting_citations}

\begin{figure}[h!]
        \centering\includegraphics[width=0.9\hsize]{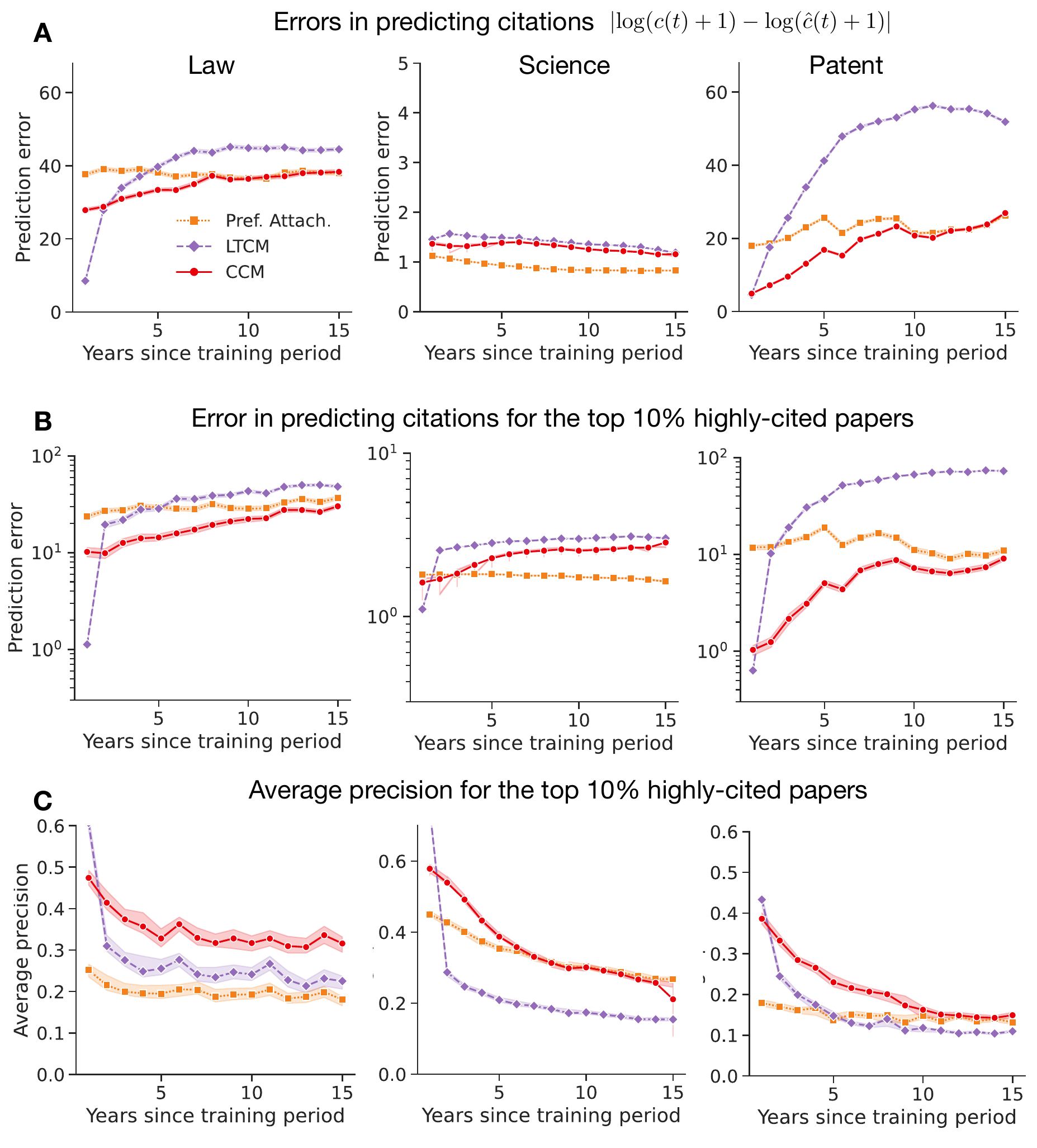}
        \caption{
                Prediction of citation counts for individual publications published in 2000 using initial citation trajectories of 5 years.
                {\bf A}. Citation prediction errors for the CCM, PA, and LTCM over time.
                The error is calculated by $\hat c(t)$, i.e., $|\log(c(t) + 1) - \log(\hat c(t) + 1)|$, where $c(t)$ and $\hat c(t)$ are the actual and predicted citations in year $t$ from the unseen publications published after the training window, respectively. 
                The 95\% confidence intervals are estimated by bootstrapping.
                {\bf B}. Citation prediction errors for the top-10\% highly-cited publications.
                The CCM has the lowest prediction errors overall, especially for young publications and highly cited publications.
                {\bf C}. Precision for identifying the top-10\% highly-cited publications. 
                A precision of 0.1 corresponds to random guessing.
        }
        \label{fig:prediction}
\end{figure}

The CCM's ability to \emph{simulate} the evolution of a whole knowledge system raises another interesting possibility:
Can this collective perspective also be used to forecast future citations?
Intuitively, a publication situated in a fast-growing niche may attract more citations than a publication in an empty space.
Estimating a publication's ``location'' in knowledge space may thus allow us to better predict future citations for all publications, including sleeping beauties.

We test this hypothesis by forecasting the citation trajectories using models trained on five years of historical citation trajectories.
Specifically, we train the PAM, LTCM, and CCM on citation trajectories from 2000 that received at least 15 citations by 2005.
We use the trained models to predict the counts of citations from unseen publications from 2006 to 2020 and compare them against the actual counts of citations from those unseen publications.
We perform the same analysis for publications published in 1990 and obtain qualitatively similar results (see SI Appendix~\ref{SI:sensitivity_analysis_prediction_cohort}).
We repeat the experiment five times using different random seeds.  

In each simulated year, $n_t$ new publications are produced, with each publication $i$ citing $m_i$ references.
For the sake of simplicity, we use the actual values of $n_t$ and $m_i$ in the test set.
We compute the prediction error by the logarithmic difference between the actual and predicted (i.e., simulated) annual citations given by $\left| \log (c(t) +1) - \log (\hat c(t) + 1)\right|$, where $c(t)$ and $\hat c(t)$ are the number of actual citations and the number of predicted citations from unseen publications from 2006 and 2020, respectively.
Note that $c(t)$ and $\hat c(t)$ are not the accumulated number of citations at time $t$; they represent new citations that the publication received after the training time window (i.e., year 2006 and onward). 

Although the prediction errors tend to converge to those of baseline models (especially PAM), the CCM makes the best short-term predictions, suggesting that its simulated community dynamics, despite its random zero-knowledge nature, capture the collective dynamics (Fig.~\ref{fig:prediction}A).
However, the CCM excels at predicting the citations trajectories for top-10\% most cited publications, maintaining a sizable margin compared to the other models over time (Fig.~\ref{fig:prediction}B).
It is important to distinguish highly cited papers from sleeping beauties: although sleeping beauties are, by construction, highly cited, highly cited papers also include those that are heavily cited shortly after publication.
The CCM's superior performance also holds for even more highly cited publications (e.g., top-5\%; see SI Appendix~\ref{sec:sensitivity_analysis_prediction_percentile}).
To further validate the predictability of citations, we evaluate the citation models using an information retrieval setup.
We use the same training data as before to train the models to identify the top-10\% most cited publications among all publications that earned at least 15 citations by 2005.
The results confirm that the CCM achieves the highest average precision, further underscoring the strong prediction capacity of CCM for highly cited publications (Fig.~\ref{fig:prediction}C).
The CCM's success in predicting highly cited publications can be attributed to collective dynamics, a factor considered by the CCM but not by the PAM and LTCM.

We test if the performance of the CCM results from the larger number of parameters as compared to the LTCM.
To this end, we limit the embedding dimensionality to $K=3$, resulting in a smaller CCM that has the same number of parameters as the LTCM (SI Appendix~\ref{sec:model_size} for the number of parameters).
Even this small CCM achieves higher precision in predicting highly-cited publications than the LTCM (see SI Figure~\ref{fig:sensitivity_dimension}).

Overall, the CCM surpasses the PAM and LTCM in predicting which publications will have the most significant future impact.
More importantly, the collective dynamics, not captured by the PAM and LTCM, appear to have a substantial effect on predicting the trajectories of the most impactful publications.

\clearpage
\section{Discussion}
\label{sec:discussion}

We report that, using a newly available corpus of U.S. judicial opinions, three fundamental knowledge systems share universal large-scale citation patterns and propose a holistic model that succinctly captures three fundamental mechanisms of citation processes---relevance, cumulative advantage, and fitness---that are sufficient to reproduce these large-scale patterns.
Our results suggest that the formalization of knowledge in human knowledge systems may be driven by a small set of mechanisms.
For instance, preferential attachment may robustly emerge across all knowledge systems due to the inherent difficulty of discovering a particular piece of knowledge from a massive preexisting pool, which subsequently creates a reliance on social signals. This herding behavior may also be behind the delayed recognition phenomenon when there is an initial lack of social signals to cite an otherwise important and useful knowledge.
A large body of prior work leverages citations to understand scientific systems~\cite{teufel1999argumentative, latour2013laboratory, etzkowitz2000dynamics, price1965networks, barabasi1999} and we contribute to this literature by studying three distinct systems in a quantitative manner and offering a model that can make predictions about future citation trajectories across a diverse range of knowledge systems.

Our study distinguishes knowledge formalization from knowledge production.
While science, law, and patents employ different methods to generate new knowledge---and the individuals in these systems operate under different procedures and incentives---we find evidence that the formalization of knowledge by means of citations across the systems exhibits universality.
We further show that modeling knowledge systems as collective processes more accurately captures the dynamics of knowledge formalization.
Our model rests on three basic and well documented assumptions, namely, that publications vary in their inherent quality, that publications connected by citation links are related and tend to cluster in the space of possible topics, and that publications with past success are likely to experience future success (cumulative advantage or preferential attachment).
These mechanisms emphasize that, despite their ostensible differences and the different time spans of these datasets, the activity of collective human knowledge systems may yield convergent large-scale patterns.

We introduce the Community Citation Model (CCM), which explains empirical citation patterns through the lens of knowledge communities and collective behavior rather than individual publications.
Unlike existing models that track long-term citation dynamics on a per-publication basis, our model conceptualizes a high-dimensional knowledge space in which publications are situated.
This space lets us define the proximity between publications that underpins their relative relevance and, thus, citation likelihood.
We note that proximity in this space is related to citation likelihood rather than topical relevance, as emphasized by the legal knowledge space that is structured by courts rather than topics.
The conceptualization of a knowledge space also allows us to provide an intuitive explanation for collective phenomena like aging and delayed recognition. 
Publications ``age'' as a community leaves a niche; sleeping beauty publications are ``discovered'' when a knowledge community shifts its attention to their niche.
These phenomena, rather than being explicitly added to the model, \emph{emerge} naturally. 


We find that modeling the knowledge space as a whole allows us to better predict the future citation trajectories of publications, particularly those that will be highly cited.
This out-performance is likely due to the fact that the CCM can take into account the publication's location in space and the trajectories of existing knowledge communities.
These insights may be leveraged to identify specific areas of science, technology, and law where innovation is likely to occur, which may be used to make decisions around funding, promotion, or advocacy.

Our study leaves room for future work in several areas.
First, while scientific citations are, to some degree, international, the patent and legal datasets we employ are limited to the United States.
Obtaining this data at scale for other jurisdictions can be challenging, but an important extension of our work will include data from other countries; especially other common law jurisdictions where judges---like U.S. judges---rely on citations. 
Second, the location of publications in the CCM is inferred based on citations, although other possibilities exist. 
Future work could investigate the possibility of combining alternative knowledge-space embeddings (e.g., those based on content). 
Third, we do not distinguish between types of citations although citations can be made to, for instance, either support or disagree with the target publication~\cite{small1978cited}. 
Although we are excited by the fact that this additional information is not needed to create a robust model, adding hidden citations~\cite{merton1968social, meng2023hidden} and distinguishing between positive and negative citations would be valuable extensions. 
Relatedly, our work could be extended by considering important qualitative aspects of citation decisions such as scientific disagreement or career advancement.
In particular, this work could shed light on how participants in knowledge systems make citation decisions and the process of re-discovering older publications.
Fourth, we focus on a spherical space for the sake of simplicity, although this is not necessarily the optimal way to model the knowledge space. 
One could alternatively explore different types of geometric space, such as Euclidean space and hyperbolic space.
Specifically, hyperbolic space has the potential to generate diverse network structures with fewer embedding dimensions, including hierarchical structure, heterogeneous degree distributions, and strong clustering~\cite{boguna2021network}.
Finally, our insights related to cumulative citation-based knowledge systems could be compared to other systems, such as online news, which have been shown to exhibit different statistical citation patterns~\cite{medo2022simple}.

Yet, despite these limitations, we argue that our model provides important new insights into fundamental mechanisms that shape the web of citations in human knowledge systems.
Our results may also open up new approaches to understanding knowledge systems and incentivizing the creation of new knowledge across different systems. 
While science and technology have been well studied, we contribute by also examining legal citation patterns that have been underexplored, and we hope this will catalyze more work on the legal system. 
Law is a critical knowledge system for society and understanding how the law evolves is important to understanding social changes. Our model sets the stage for the generalized study of citations across different systems~\cite{ahmadpoor2017dual} to unveil the complex scientific, intellectual, and legal ecosystems around e.g., recent technological disruptions. 
Moreover, although our focus is on knowledge formalization via citations, our methodology could be used to delve into the processes of knowledge production itself to understand and forecast the nuanced dynamics within and across these systems, such as disruption, innovation, and structural differences between communities in law, science, and technology. 
Ultimately, we argue that the production of knowledge is a collective phenomenon, which can only be properly understood by taking a systematic and holistic approach.

\section{Methods}
\label{sec:methods}

\subsection{Data Sources}
\label{sec:data_sources}

For the legal system data, we use the Case Law Access Project Citation Graph that includes almost two million U.S. federal judicial opinions published between 1781 and 2019, and 15 million citations between them.


For scientific system data, we use the SciSciNet data~\cite{lin2023sciscinet} that contains  $134,129,188$ publications and $1,588,739,703$ citations.
The SciSciNet data is roughly 18 times larger than the second largest dataset (Patent) in our analysis, and we find that it requires substantial time to perform our analyses. Thus, we reduced the size of the dataset by excluding the publications published after 2016, which results in $58,299,343$ publications and $973,266,138$ citations. 
Additionally, we employ another science citation dataset, American Physical Society (APS) dataset, for computationally expensive analysis e.g., ablation study.

For the patent system data, we use the dataset from PatentsView containing $7,430,873$ patents and $97,196,273$ citations registered in the US Patent and Trademark Office (USPTO).
The patent citations may not be as complete as what might be expected in the scientific literature~\cite{sampat2010applicants}.
The reason for this under-citation pattern is that patent filers are incentivized to distinguish their inventions from other work in order to satisfy the legal requirements for patentability (see 35 U.S. Code § 101-103).
Although this bias poses a challenge for other bibliometric research, the different incentives for inventors make the universal trends observed in Section~\ref{sec:universal_patterns} all the more surprising and underscore that the universal trends are robust against differences between knowledge systems.

\subsection{Estimating Exponential Growth}
\label{sec:exponential_growth}

We estimate the annual growth rate, $\alpha$, of publications, citations, and references (see Fig.~\ref{fig:citation-dynamics}A and Fig.~\ref{fig:growth}) by a Poisson model with mean
$
{\mathbb E}[n_y \vert y] = \exp\left(\alpha y\right),
$
where $n_y$ is the number of publications, citations, or references in year $y$.
The parameter $\alpha$ is obtained by maximizing the log-likelihood using the \texttt{scikit-learn} package~\cite{scikit-learn}.

\subsection{Estimating Preferential Attachment and Recency}
\label{sec:preferential_attachment}


We use the \texttt{powerlaw} package~\cite{Alstott2014-bm} to perform statistical tests for the power-law degree distribution and to fit $\alpha$ using log-likelihood.
As shown in earlier studies, we confirm that the number of new citations is proportional to the number of accumulated citations across the three datasets (see Fig.~\ref{fig:citation-dynamics}C).
For visual clarity, we bin the accumulated citations with bins spaced evenly on a log scale.

The recency bias counteracts preferential attachment and the inflation of citations.
Preferential attachment increases the citations of highly-cited publications, while citation inflation increases the average number of annual citations, thus increasing the likelihood that a publication recieves new citations.
We control for these two effects which counteract aging to identify the empirical aging function $f^{\text{emp}}$ as follows:
First, we control for the effect of preferential attachment by focusing on the time when a publication $j$ receives $x$ citations.
We then calculate the time, $\Delta t_{j,x+1}$, between publication until the next citation (the $(x+1)$-th citation). 
Second, we control for the effect of citation inflation by normalizing the total number of citations, $C_j$,generated in the year of the $(x+1)$-th citation for publication $j$, i.e.,  
\begin{align}
f^{\text{emp}}(\Delta t \vert x) = \frac{1}{R}\sum_{j=1}^N \frac{1}{C_j} \mathbb{1}\left\{\Delta t_{j,x+1} = \Delta t \right\}, 
\end{align}
where $N$ is the number of publications, and $R$ is a normalization constant to ensure that $f^{\text{emp}}$ represents the same probability as in the aging function of the LTCM, i.e., $R = \sum_{\Delta t}f^{\text{emp}}(\Delta t \vert x)$.

\subsection{Estimating Delayed Recognition}
\label{sec:delayed_recognition}

We follow an existing geometric approach to quantify delayed recognition (``sleeping beauties'')~\cite{Ke2015defining}:
For a given publication, $c_y$ is the number of citations the publication has received in year $y$.
We rescale the publication's release year to 0, such that the point $(0,c_0)$ denotes the first available point in the time-citation plane.
We denote by $y_m$ the year in which the publication receives the maximum number of citations.
Then, we define a linear reference line $l_y$ as the straight line that connects $(0,c_0)$ and $(y_m, c_{y_m})$.
The publication's sleeping beautify coefficient $B$ is subsequently defined as
\begin{align}
        B = \sum_{t=0}^{t=m} \frac{l_t - c_t}{\max(1,c_t)},
\end{align}
which can be interpreted as the area under the curve between the reference line $l_t$ and the citation history line $(y, c_y)_{y=0}^{y_m}$.
The larger $B$, the larger the deviation between the two lines, and hence the more suddenly the maximum citation was achieved.
A coefficient of $B=0$ corresponds to gradual linear citation growth.
Situations where $B$ is negative present the opposite: publications that initially attract a lot of attention and subsequently phase out.
We calculate $B$ in units of years and find a remarkably similar distribution across all three knowledge systems (Fig.~\ref{fig:citation-dynamics}E).


\subsection{Fitting the CCM}
\label{sec:CCM_fitting}

We fit the CCM using a two-step process.
In the first step, we estimate parameters that determine the citation probabilities
$
\theta \equiv \left( \left\{ \vec{u}_j \right\}, \left\{ \eta_j \right\}, \kappa \right)
$ via Eq.~\eqref{eq:CCM}.
In the second step, we estimate the scaling $\lambda$ of the publication placement process via Eq.~\eqref{eq:adj_prod} while keeping $\theta$ fixed.

In the first step, we estimate the parameters
$
\theta = \left( \left\{ \vec{u}_j \right\}, \left\{ \eta_j \right\}, \kappa \right)
$
from Eq.~\eqref{eq:CCM} given the publications' citation histories.
Maximizing the likelihood via gradient ascent is computationally infeasible because the normalization constant $Z(t_j)$ requires summation across all publications.
We thus use Noise Contrastive Estimation (NCE)~\cite{dyer2014notes,gutmann2010noise}.
The key idea underlying NCE is that it creates an easy-to-maximize surrogate function that has a maximum at the same parameter value as the original log-likelihood function of interest
(see Appendix \ref{SI:model_fitting} for details).
We used NCE to simultaneously estimate the parameters, $\theta =\left( \left\{ \vec{u}_j \right\}, \left\{ \eta_j \right\}, \kappa \right)$.
While we set the embedding dimension $K$ to 128, we note that qualitatively similar results are obtained with different values for $K$ so long as $K$ is not too small ($K \geqslant 64$; see SI Appendix~\ref{sec:sensitivity_analysis_hyperparameters}).

In the second step, we estimate the scaling factor $\lambda$ that governs the distribution of publications in the embedding space, while keeping the previously estimated $\theta$---which governs citations between publications---fixed.
The probability that a new publication published in year $t$ finds its place at location $\vec{u}$ is given by 
\begin{align}
 P(\vec{u} \given \{\vec{u_j};\; t_j < t\}, \lambda )=\frac{1}{N_{t}}\sum_{j,\; t_j < t_i} P(\vec{u} \given \vec{u_j}), \label{eq:emalgorithm}
\end{align}
where $N_t$ is the number of existing publications that are citable for publication $\vec{u}$.
Equation~\eqref{eq:emalgorithm} is a mixture of von Mises distributions, each centered on a previous publication in the embedding space.
We employ the expectation-maximization (EM) algorithm to fit the mixture model to estimate $\lambda$.
Specifically, we fit the mixture model by maximizing the log-likelihood given by 
\begin{align}
J := \sum_{i=1}^N \log P(\vec{u_i} \given \{\vec{u_j};\; t_j < t_i\}, \lambda )=\sum_{i=1}^N \log \left[ \frac{1}{N_{t_i}}\sum_{j,\; t_j < t_i} P(\vec{u_i} \given \vec{u_j})\right]. \label{eq:emalgorithm_log_likelihood}
\end{align}
Maximizing $J$ is a non-trivial task because the log of a summation is not concave~\cite{bishop2006pattern}.
As commonly done in EM algorithms, we thus introduce a latent variable $q_{ij}$ representing the probability that publication $i$ is drawn from the von Mises distribution centered at $j$, where $\sum_j q_{ij} = 1$ for all $i$.
Then, we rewrite Eq.~\eqref{eq:emalgorithm_log_likelihood} as 
\begin{align}
J&=\sum_{i=1}^N \log \left[ \sum_{j,\; t_j < t_i} q_{ij}\dfrac{\frac{1}{N_{t_i}}P(\vec{u_i} \given \vec{u_j})}{q_{ij}}\right] =\sum_{i=1}^N \log \left( \sum_{j,\; t_j < t_i} {\mathbb E}_{q_{ij}}\left[ \dfrac{\frac{1}{N_{t_i}}P(\vec{u_i} \given \vec{u_j})}{q_{ij}}\right]\right), \label{eq:emalgorithm_log_likelihood2}
\end{align}
where $\mathbb{E}_{q_{ij}}[\cdot]$ represents the weighted average over $j$, with weights $\left\{q_{ij} \right\}_j$. 
By using Jensen’s inequality (i.e., $\log {\mathbb E}[h(x)] \geq {\mathbb E}[\log h(x)] $ for a concave function $h$), we obtain 
\begin{align}
J \geq J' &=\sum_{i=1}^N \sum_{j,\; t_j < t_i} q_{ij}\log \dfrac{\frac{1}{N_{t_i}}P(\vec{u_i} \given \vec{u_j})}{q_{ij}}.\label{eq:emalgorithm_log_likelihood3}
\end{align}
Equation~\eqref{eq:emalgorithm_log_likelihood3} suggests that $J$ can be maximized by maximizing $J'$ which does not involve the log of the summation. 
Maximizing $J'$ is a constrained maximization problem under constraint $\sum_{j} q_{ij} = 1$ for all $i$. 
The Lagrangian function ${\cal L}$ for this constrained maximization problem is given by
\begin{align}
{\cal L}: = \sum_{i=1}^N \sum_{j,\; t_j < t_i} q_{ij}\log \dfrac{\frac{1}{N_{t_i}}P(\vec{u_i} \given \vec{u_j})}{q_{ij}} + \sum_{i} \alpha_i \left(\sum_{j} q_{ij} -1\right),
\end{align}
where $\alpha_i$ is the Lagrange multiplier. 
By solving $\partial {\cal L} / \partial q_{ij} = 0$, we have 
\begin{align}
\log \dfrac{\frac{1}{N_{t_i}}P(\vec{u_i} \given \vec{u_j})}{q_{ij}} -1+ \alpha_i = 0 \iff q_{ij} = \frac{1}{N_{t_i}}P(\vec{u_i} \given \vec{u_j})\exp(\alpha_i -1).
\end{align}
Because $\sum_j q_{ij}=1$, we have
\begin{align}
q_{ij} = \dfrac{P(\vec{u_i} \given \vec{u_j})}{\sum_{\ell} P(\vec{u_i} \given \vec{u_\ell})}. \label{eq:q_def}
\end{align}

Let us consider optimizing the other parameter, $\lambda$, using a gradient descent algorithm. 
The gradient is given by
\begin{align}
\frac{\partial {\cal L} }{\partial \lambda} = 
\sum_{i}\sum_{j,\; t_j < t_i} q_{ij} \frac{\partial}{\partial \lambda}\left[ \log P(\vec{u_i} \given \vec{u_j})\right]. \label{eq:equation_for_lambda}
\end{align}
Computing this derivative is challenging because the von Mises distribution, $P(\vec{u_i} \given \vec{u_j})$, involves the analytically intractable normalization constant that contains $\lambda$. 
To circumvent this problem, we take advantage of the relationship between von Mises distributions and Gaussian distributions.
Namely, a Gaussian distribution on a $K$-dimensional space centered at $\vec{u}$ with variance $\lambda^{-1}\mat{I}$ is given by
\begin{align}
{\cal N}(\vec{x} \given \vec{u},\lambda^{-1}) = 
\left(\frac{\lambda}{2\pi}\right)^{\frac{K}{2}} \exp\left[ -\frac{\lambda}{2}(\vec{x} - \vec{u})^\top (\vec{x} - \vec{u})  \right].
\end{align}
By expanding the square term, we obtain
\begin{align}
{\cal N}(\vec{x} \given \vec{u}, \lambda^{-1}) &= \left(\frac{\lambda}{2\pi}\right)^{\frac{K}{2}} \exp \left[-\frac{\lambda}{2}\vec{x}^\top \vec{x}-\frac{\lambda}{2}\vec{u}^\top \vec{u}\right] \cdot \exp \left[ \lambda\vec{x}^\top \vec{u}\right]. \label{eq:gaussian_vonmises}
\end{align}
Note that $\vec{u}^\top \vec{u}$ is constant since the vector length is fixed.
If random vector $\vec{x}$ has a similar vector length, Eq.~\eqref{eq:gaussian_vonmises} can be well approximated by the von Mises distribution with $d(\vec{u}, \vec{x}) = 1-\vec{u}^\top \vec{x}$.
This assumption holds mostly true for high-dimensional vectors because (i) the vector length follows a $\chi^2$-distribution, (ii) the standard deviation of the vector length scales with $\sqrt{2K}$ while the mean scales with $K$, and thus, (iii) the relative variation of the vector length to its mean, $\lim_{K\rightarrow \infty} \sqrt{2K}/K \rightarrow 0$, i.e., the vectors generated by a Gaussian distribution have almost the same length relative to their mean.
This approximation enables us to obtain a closed-form expression 
\begin{align}
\frac{\partial}{\partial \lambda}\left[ \log P(\vec{u_i} \given \vec{u_j})\right] \simeq \frac{K}{2\lambda} - (1-\vec{u_i}^\top \vec{u_j}).
\label{eq:logp_approx}
\end{align}

Based on Eqs.~\eqref{eq:q_def} and ~\eqref{eq:logp_approx}, we now define our EM algorithm. 
The EM algorithm updates $q_{ij}$ and $\lambda$ by maximizing $J'(\lambda, \{q_{ij}\},\{\vec{u}_{i}\})$ with respect to either $q_{ij}$ and $\lambda$ while fixing the other.
The maximization can be done analytically for $q_{ij}$ by solving $\partial J'(\lambda, \{q_{ij}\},\{\vec{u}_{i}\})/\partial q_{ij} = 0$.
Furthermore, the maximization with respect to $\lambda$ can be achieved by solving $\partial J'(\lambda, \{q_{ij}\},\{\vec{u}_{i}\})/\partial \lambda = 0$.
This results in the following recursive updating rules:
\begin{align}
q_{ij} ^{(\ell + 1)}
&=
\frac{\exp[-\lambda^{(\ell)} d( \vec{u}_i, \vec{v}_j)]}{\sum_{j',\; t_{j'} \leq t}\exp[-\lambda^{(\ell)} d( \vec{u}_i, \vec{v}_j')]}, \label{eq:update_q} \\
\lambda^{(\ell + 1)} &
=
\lambda^{(\ell)} + \epsilon \sum_{i}\sum_{j, t_j < t_i} q_{ij} ^{(\ell + 1)}\left[\frac{K}{2\lambda^{(\ell)}} - d(\vec{u_i}, \vec{u_j})\right], \label{eq:update_lambda}
\end{align}
where $\epsilon$ is a learning rate.
We determine the learning rate $\epsilon$ by using the adaptive moment estimation (ADAM), with hyperparameters set to those in the original paper~\cite{kingma2014adam}.

\subsection{Predicting Citation Trajectories}
\label{sec:CCM_predictions}

We predict future citations by using publications' initial citation history.
More specifically, we train the CCM with all citations from publications published up to time $t_{\text{train}}$.
Then, we generate the network for $t> t_{\text{train}}$ in the following two steps.
First, we generate the locations of the new publications according to Eq.~\eqref{eq:adj_prod}.
Note that we do not provide citations to publications at this time.
We determine the number of new publications and the number of references for each new publication based on the actual number of publications and references in the original networks. 
Furthermore, we assign the fitness value of a new publication by randomly sampling a fitness value of the publications in the training data.

Second, we generate the citations from the new publications according to Eq.~\eqref{eq:CCM} based on the generated publication embeddings.
This often involves the computation of the citation probability between every new publication and millions of existing publications, which is computationally expensive.
While there are heuristics to accelerate sampling at the expense of accuracy, we find that exact sampling yields the best citation prediction. 
Thus, we compute the exact probabilities of citations from each new publication to every existing publications. 
We employ the Efraimidis-Spirakis algorithm~\cite{efraimidis2006weighted} to perform the sampling without replacement.

\subsection{Long-Term Citation Model}
\label{sec:long_term_citation_model}

We fit the LTCM based on maximum likelihood estimation~\cite{shenModelingPredictingPopularity2014}.
We use the ADAM gradient descent algorithm~\cite{kingma2014adam} to maximize the likelihood of the LTCM with respect to $(\eta_j, \mu_j, \sigma_j)$ in Eq.~\eqref{eq:LTCM} for individual paper $j$.
The LTCM produces a sequence of citation events for individual publications but does not produce the citation network.
Thus, we construct the citation network as follows:
First, we compute the expected number of citation events for individual papers based on Eq.~(8) in Ref.~\cite{shenModelingPredictingPopularity2014}.
The computed number of citation events are real values, and we round them to the nearest integer. 
For each citation event, we add an edge from the publication to a randomly selected publication that is published at the time of citation, thus generating a citation network.


\subsection{Data Availability Statement}
\label{sec:data_availability}
Three datasets were used in this study. The USPTO data and U.S. Common Law Data are both available via FigShare~\url{https://figshare.com/projects/Uncovering\_the\_universal\_dynamics\_of\_citation\_systems\_From\_science\_of\_science\_to\_law\_of\_law\_and\_patterns\_of\_patents/202335}. 
The SciSciNet data can be obtained from \url{https://doi.org/10.6084/m9.figshare.c.6076908.v1}, and the code to process this data is available in our code repository. 

\subsection{Code Availability Statement}
\label{sec:code_availability}
The code to reproduce the results is available at \url{https://github.com/skojaku/community_citation_model}.

{
\footnotesize
\bibliography{bibliography.bib}

\begin{thebibliography}{10}

\bibitem{small1978cited}
Henry~G. Small.
\newblock Cited {{Documents}} as {{Concept Symbols}}.
\newblock 8(3):327--340.

\bibitem{latour2013laboratory}
Bruno Latour and Steve Woolgar.
\newblock {\em Laboratory life: The construction of scientific facts}.
\newblock Princeton university press, 2013.

\bibitem{Price1976}
Derek John de~Solla Price.
\newblock A general theory of bibliometric and other cumulative advantage
  processes.
\newblock {\em Journal of the American society for Information science},
  27(5):292--306, 1976.

\bibitem{garfield1955citation}
Eugene Garfield.
\newblock Citation indexes for science: A new dimension in documentation
  through association of ideas.
\newblock {\em Science}, 122(3159):108--111, 1955.

\bibitem{price1965networks}
Derek John de~Solla Price.
\newblock Networks of scientific papers: The pattern of bibliographic
  references indicates the nature of the scientific research front.
\newblock {\em Science}, 149(3683):510--515, 1965.

\bibitem{shapiro1992origins}
Fred~R Shapiro.
\newblock Origins of bibliometrics, citation indexing, and citation analysis:
  The neglected legal literature.
\newblock {\em Journal of the American Society for Information Science},
  43(5):337--339, 1992.

\bibitem{price1963little}
Derek John de~Solla Price.
\newblock {\em Little science, big science}.
\newblock Columbia University Press, 1963.

\bibitem{barabasi1999}
Albert-L\'{a}szl\'{o} Barab\'{a}si and R\'{e}ka Albert.
\newblock Emergence of scaling in random networks.
\newblock {\em Science}, 286(5439):509--512, 1999.

\bibitem{candia2019universal}
Cristian Candia, C~Jara-Figueroa, Carlos Rodriguez-Sickert,
  Albert-L{\'a}szl{\'o} Barab{\'a}si, and C{\'e}sar~A Hidalgo.
\newblock The universal decay of collective memory and attention.
\newblock {\em {Nature Human Behaviour}}, 3(1):82--91, 2019.

\bibitem{candia2020rising}
Cristian Candia and Brian Uzzi.
\newblock The rising of collective forgetting and cultural selectivity in
  inventors and physicists communities.
\newblock {\em arXiv preprint arXiv:2008.06592}, 2020.

\bibitem{radicchi2008universality}
Filippo Radicchi, Santo Fortunato, and Claudio Castellano.
\newblock Universality of citation distributions: toward an objective measure
  of scientific impact.
\newblock {\em Proceedings of the National Academy of Sciences},
  105(45):17268--17272, November 2008.

\bibitem{milojevic2015quantifying}
Sta{\v{s}}a Milojevi{\'c}.
\newblock Quantifying the cognitive extent of science.
\newblock {\em Journal of Informetrics}, 9(4):962--973, 2015.

\bibitem{Ke2015defining}
Qing Ke, Emilio Ferrara, Filippo Radicchi, and Alessandro Flammini.
\newblock Defining and identifying sleeping beauties in science.
\newblock {\em Proceedings of the National Academy of Sciences},
  112(24):7426--7431, 2015.

\bibitem{fortunato2018science}
Santo Fortunato, Carl~T Bergstrom, Katy B{\"o}rner, James~A Evans, Dirk
  Helbing, Sta{\v{s}}a Milojevi{\'c}, Alexander~M Petersen, Filippo Radicchi,
  Roberta Sinatra, Brian Uzzi, et~al.
\newblock Science of science.
\newblock {\em Science}, 359(6379), 2018.

\bibitem{wang2013quantifying}
Dashun Wang, Chaoming Song, and Albert-L{\'a}szl{\'o} Barab{\'a}si.
\newblock Quantifying long-term scientific impact.
\newblock {\em Science}, 342(6154):127--132, 2013.

\bibitem{Sinatra2016-oi}
Roberta Sinatra, Dashun Wang, Pierre Deville, Chaoming Song, and
  Albert-L{\'a}szl{\'o} Barab{\'a}si.
\newblock Quantifying the evolution of individual scientific impact.
\newblock {\em Science}, 354(6312):aaf5239, 2016.

\bibitem{garfield1957breaking}
Eugene Garfield.
\newblock Breaking the subject index barrier: A citation index for chemical
  patents.
\newblock {\em Journal of the Patent Office Society}, 39(5):583, 1957.

\bibitem{small1973co}
Henry Small.
\newblock Co-citation in the scientific literature: A new measure of the
  relationship between two documents.
\newblock {\em Journal of the American Society for information Science},
  24(4):265--269, 1973.

\bibitem{small1985clustering}
Henry Small, Ernest Sweeney, and Edward Greenlee.
\newblock Clustering the science citation index using co-citations. ii. mapping
  science.
\newblock {\em Scientometrics}, 8:321--340, 1985.

\bibitem{kessler1963bibliographic}
Maxwell~Mirton Kessler.
\newblock Bibliographic coupling between scientific papers.
\newblock {\em American documentation}, 14(1):10--25, 1963.

\bibitem{teufel1999argumentative}
Simone Teufel.
\newblock {\em Argumentative Zoning: Information Extraction from Scientific
  Text}.
\newblock PhD thesis, University of Edinburgh, Edinburgh, UK, 1999.

\bibitem{merton1973sociology}
Robert~K Merton.
\newblock The sociology of science: Theoretical and empirical investigations.
\newblock {\em The University of Chicago}, 1973.

\bibitem{van2004sleeping}
Anthony~FJ Van~Raan.
\newblock Sleeping beauties in science.
\newblock {\em Scientometrics}, 59:467--472, 2004.

\bibitem{stent1972prematurity}
Gunther~S Stent.
\newblock Prematurity and uniqueness in scientific discovery.
\newblock {\em Scientific American}, 227(6):84--93, 1972.

\bibitem{zuckerman1980indicators}
Harriet Zuckerman and Roberta Miller.
\newblock Indicators of science: Notes and queries.
\newblock {\em Scientometrics}, 2(5-6):347--353, 1980.

\bibitem{merton1968social}
Robert~King Merton.
\newblock {\em Social theory and social structure}.
\newblock Simon and Schuster, 1968.

\bibitem{fathally2009systemes}
Jabeur Fathally, Nicola Mariani, Louis Perret, and Alain-Fran{\c{c}}ois Bisson.
\newblock {\em Les syst{\`e}mes juridiques dans le monde}.
\newblock Collection Bleue. Wilson \& Lafleur, 2 edition, 2009.

\bibitem{park2023papers}
Michael Park, Erin Leahey, and Russell~J Funk.
\newblock Papers and patents are becoming less disruptive over time.
\newblock {\em Nature}, 613(7942):138--144, 2023.

\bibitem{funk2017dynamic}
Russell~J Funk and Jason Owen-Smith.
\newblock A dynamic network measure of technological change.
\newblock {\em Management Science}, 63(3):791--817, 2017.

\bibitem{marx2020reliance}
Matt Marx and Aaron Fuegi.
\newblock Reliance on science: Worldwide front-page patent citations to
  scientific articles.
\newblock {\em Strategic Management Journal}, 41(9):1572--1594, 2020.

\bibitem{jones2009burden}
Benjamin~F Jones.
\newblock The burden of knowledge and the ``death of the renaissance man'': Is
  innovation getting harder?
\newblock {\em The Review of Economic Studies}, 76(1):283--317, 2009.

\bibitem{krapivsky2005network}
Pavel~L Krapivsky and Sidney Redner.
\newblock Network growth by copying.
\newblock {\em Physical Review E—Statistical, Nonlinear, and Soft Matter
  Physics}, 71(3):036118, 2005.

\bibitem{Medo2011-ko}
Mat{\'u}{\v s} Medo, Giulio Cimini, and Stanislao Gualdi.
\newblock Temporal effects in the growth of networks.
\newblock {\em Physical Review Letters}, 107(23):238701, December 2011.

\bibitem{Ke2015}
Qing Ke.
\newblock Qualifying exam report: On the universal rescaled citation dynamics
  of individual papers.
\newblock {\em Indiana University}, pages 1--13, 2015.

\bibitem{shenModelingPredictingPopularity2014}
Huawei Shen, Dashun Wang, Chaoming Song, and Albert-L{\'a}szl{\'o}
  Barab{\'a}si.
\newblock Modeling and predicting popularity dynamics via reinforced poisson
  processes.
\newblock {\em Proceedings of the AAAI Conference on Artificial Intelligence},
  28(1), Jun. 2014.

\bibitem{barabasi2018formula}
Albert-László Barab\'{a}si.
\newblock {\em The formula: {The} universal laws of success}.
\newblock Little Brown and Company, 2018.

\bibitem{upham2010emerging}
S~Upham and Henry Small.
\newblock Emerging research fronts in science and technology: patterns of new
  knowledge development.
\newblock {\em Scientometrics}, 83(1):15--38, 2010.

\bibitem{boyack2005mapping}
Kevin~W Boyack, Richard Klavans, and Katy B{\"o}rner.
\newblock Mapping the backbone of science.
\newblock {\em Scientometrics}, 64(3):351--374, 2005.

\bibitem{borner2003visualizing}
Katy B{\"o}rner, Chaomei Chen, and Kevin~W Boyack.
\newblock Visualizing knowledge domains.
\newblock {\em Annual review of information science and technology},
  37(1):179--255, 2003.

\bibitem{etzkowitz2000dynamics}
Henry Etzkowitz and Loet Leydesdorff.
\newblock The dynamics of innovation: from national systems and “mode 2” to
  a triple helix of university--industry--government relations.
\newblock {\em Research policy}, 29(2):109--123, 2000.

\bibitem{meng2023hidden}
Xiangyi Meng, Onur Varol, and Albert-L{\'a}szl{\'o} Barab{\'a}si.
\newblock Hidden citations obscure true impact in science.
\newblock {\em arXiv preprint arXiv:2310.16181}, 2023.

\bibitem{boguna2021network}
Marian Boguna, Ivan Bonamassa, Manlio De~Domenico, Shlomo Havlin, Dmitri
  Krioukov, and M~{\'A}ngeles Serrano.
\newblock Network geometry.
\newblock {\em Nature Reviews Physics}, 3(2):114--135, 2021.

\bibitem{medo2022simple}
Mat{\'u}{\v{s}} Medo, Manuel~S Mariani, and Linyuan L{\"u}.
\newblock The simple regularities in the dynamics of online news impact.
\newblock {\em Journal of computational social science}, pages 1--18, 2022.

\bibitem{ahmadpoor2017dual}
Mohammad Ahmadpoor and Benjamin~F Jones.
\newblock The dual frontier: Patented inventions and prior scientific advance.
\newblock {\em Science}, 357(6351):583--587, 2017.

\bibitem{lin2023sciscinet}
Zihang Lin, Yian Yin, Lu~Liu, and Dashun Wang.
\newblock Sciscinet: A large-scale open data lake for the science of science
  research.
\newblock {\em Scientific Data}, 10(1):315, 2023.

\bibitem{sampat2010applicants}
Bhaven~N Sampat.
\newblock When do applicants search for prior art?
\newblock {\em The Journal of Law and Economics}, 53(2):399--416, 2010.

\bibitem{scikit-learn}
F.~Pedregosa, G.~Varoquaux, A.~Gramfort, V.~Michel, B.~Thirion, O.~Grisel,
  M.~Blondel, P.~Prettenhofer, R.~Weiss, V.~Dubourg, J.~Vanderplas, A.~Passos,
  D.~Cournapeau, M.~Brucher, M.~Perrot, and E.~Duchesnay.
\newblock Scikit-learn: Machine learning in {P}ython.
\newblock {\em Journal of Machine Learning Research}, 12:2825--2830, 2011.

\bibitem{Alstott2014-bm}
Jeff Alstott, Ed~Bullmore, and Dietmar Plenz.
\newblock Powerlaw: a python package for analysis of heavy-tailed
  distributions.
\newblock {\em PLoS One}, 9(1):e85777, January 2014.

\bibitem{dyer2014notes}
Chris Dyer.
\newblock Notes on noise contrastive estimation and negative sampling.
\newblock {\em arXiv preprint arXiv:1410.8251}, 2014.

\bibitem{gutmann2010noise}
Michael Gutmann and Aapo Hyv{\"a}rinen.
\newblock Noise-contrastive estimation: A new estimation principle for
  unnormalized statistical models.
\newblock In {\em International Conference on Artificial Intelligence and
  Statistics}, pages 297--304, 2010.

\bibitem{bishop2006pattern}
Christopher~M Bishop and Nasser~M Nasrabadi.
\newblock {\em Pattern recognition and machine learning}, volume~4.
\newblock Springer, 2006.

\bibitem{kingma2014adam}
Diederik~P Kingma and Jimmy Ba.
\newblock Adam: A method for stochastic optimization.
\newblock {\em ICLR 2017}, 2014.

\bibitem{efraimidis2006weighted}
Pavlos~S Efraimidis and Paul~G Spirakis.
\newblock Weighted random sampling with a reservoir.
\newblock {\em Information Processing Letters}, 97(5):181--185, 2006.

\bibitem{kmiec2004origin}
Keenan~D Kmiec.
\newblock The origin and current meanings of judicial activism.
\newblock {\em California Law Review}, 92:1441, 2004.

\end{thebibliography}
}

\section*{Acknowledgements}
We thank Dr.~Staša Milojević, Dr.~Alessandro Flammini, Dr.~Filippo Menczer, and Dr.~Dashun Wang for their helpful feedback.
S.K. and Y.Y.A. are supported by the Air Force Office of Scientific Research under award number FA9550-19-1-0391. 
Y.Y.A. is in part supported by National Science Foundation under Grant No.~2404109.
S.K. and Y.Y.A. acknowledge the computing resources at Indiana University and thank NVIDIA Corporation for their GPU resources.
E.M. acknowledges support by the National Science Foundation under Grant No.~2218748.

\section*{Author Contributions}

R.M and Y.Y.A. conceived the project. S.K, R.M., and Y.Y.A. designed the analysis and the model with contributions from S.L., E.M., and A.P. S.K. led the technical implementation with contributions from all authors. All authors discussed the results. R.M., S.K., S.L., and Y.Y.A. wrote the manuscript with input from all authors.

\section*{Competing Interests}
The authors declare no competing interests.

\newpage
\appendix

\begin{center}
    \section*{Supplementary Information\\
             \large Uncovering the universal dynamics of citation systems: From science of science to law of law and patterns of patents \\
    }
\end{center}

\section{U.S. Common Law in a Nutshell}
\label{SI:us_common_law}

Broadly speaking, the U.S. federal judiciary consists of 94 district courts, 13 appellate courts and the U.S. supreme court\footnote{There are several other specialized courts which will not be examined here.}.
These courts are organized into 12 regional circuits, each of which consists of an appellate court and a set of district courts.
All federal judges are appointed by a U.S. President, confirmed by the Senate, and serve a lifetime term (removal is possible only under exceptional circumstances).
All these courts are governed by the fundamental principle that judges are bound by precedent, a doctrine known as \textit{stare decisis} (SD) which operates on two levels:
vertically and horizontally.
\textit{Vertical stare decisis} (VSD) refers to the principle that a lower court is bound by precedent from a higher court,
e.g., all federal appellate courts are bound by precedent from the U.S. Supreme Court.
\textit{Horizontal stare decisis} (HSD) refers to the principle that a court is bound by its own decisions,
e.g., a given appellate court should respect its own prior decisions.
While a lower court cannot generally ignore VSD, under certain circumstances a court may depart from HSD,
although judges tend to distinguish a given case from precedent rather than violating SD \cite{kmiec2004origin}.
The discussion above focuses on binding precedent, which must be followed, however, U.S. courts may also consider persuasive precedent,
which is not legally binding, such as opinions from lower courts, dissenting opinions, or decisions from other jurisdictions.

U.S judges are generalists, in contrast to scientists and inventors, are and they must rule on whatever cases are brought before them as a result of human processes which they cannot control.
Judges are generally assigned cases at random and do not usually specialize (although some judges may become specialists by virtue of regional differences that affect the composition of their caseloads).
Despite the unique common law rules that govern judicial citations, we find that judicial citations obey remarkably similar dynamics as scientific citations.
\clearpage
\newpage

\section{Comparing Knowledge Systems}
\label{SI:compare_knowledge_systems}

\begin{singlespace}
\begin{table}[!ht]
    \centering
    \footnotesize
    \begin{tabular}{|p{0.12\linewidth}|p{0.18\linewidth}|p{0.2\linewidth}|p{0.2\linewidth}|p{0.2\linewidth}|}
    \hline
        \textbf{Mechanism} & \textbf{Procedure} & \textbf{Law} & \textbf{Science}  & \textbf{Patents}  \\
        \hline
        \hline
        Filtering & Participation Criteria & Licensed as attorney, appointed by President & Anyone in principle, Training normally required  & Anyone in principle, Training normally required\\ \hline
        Freedom & Choice of Subject & Randomly assigned & Authors specialized, driven by personal interest & Inventors specialized, driven by personal and commercial interest  \\ \hline
        Freedom & Collaboration & Rare & Common  & Common\\
        \hline
        Freedom & Method of inquiry & Legal research, Adversarial system & Experimental method, hypothesis driven, measurement & Often experimental method, hypothesis driven, measurement  \\
        \hline
        Freedom & Time frame & Fast & Variable & Slow  \\
        \hline
        Freedom & Investment/Funding requirements & Low investment (for the author!) & Medium to high investment & High investment \\
        \hline
        Filtering & Publication process & Unilateral & Peer Review & USPTO  \\
        \hline
        Other & Discovery process & LexisNexis, Westlaw and other legal research platforms & Google Scholar and other scientific repositories & Patent research platforms and scientific repositories, but may choose not to cite \\
        \hline
        Other & Use of assistants & High: Clerks & High: RAs & Varies \\
        \hline
        Other & Hierarchy & Binding and strict & Loose & May strategically avoid citing the most successful existing patents (``inverted hierarchy'')\\
        \hline
        Other & Number of authors & Low & High & Medium \\
        \hline
        Other & Growth of number of authors & Constant & High & High \\
        \hline
        Other & Importance of geography & High (law varies from place to place) & Low (Science is similar everywhere) & Medium (patent granted in specific places) \\
        \hline
    \end{tabular}
\end{table}
\end{singlespace}

\section{Sleeping Beauties}
\label{SI:SB}

Sleeping Beauties are publications that initially receive very few citations, but experience a sudden burst of citations after some time, often as the result of a re-discovery of a body of knowledge.
Here, we present three representative examples for Sleeping Beauties from law, science, and patents.

In law, an example for a Sleeping Beauty is \textit{United States v. Hooton}, 693 F.2d 857 (1982).
The case involves an appeal to the Ninth Circuit Court of Appeals.
The court characterizes the appellant's appeal as insubstantial and dismisses it.
After the opinion is published in 1982, it is cited only 11 times in the subsequent 20 years.
In 2002, the Ninth Circuit cites \textit{Hooton} to dismiss an appeal stating that ``the question raised in this appeal is so insubstantial {[}...{]} as not to require further argument.''
Following this rediscovery, \textit{Hooton} receives over 1,000 citations in a span of five years, and serves as a basis by which to dismiss an ``insubstantial'' appeal with a very short opinion.

In science, a striking example of a Sleeping Beauty is the 1935 article published by Albert Einstein, Boris Podolsky and Nathan Rosen.
The paper, titled ``Can quantum-mechanical description of physical reality be considered complete?'' received little attention for almost 70 years before experiencing a surge of citations~\cite{zuckerman1980indicators, Ke2015defining}.

In patents, we identify ``Thermal deferred action battery'' (US 3972734A) as an example of a prototypical Sleeping Beauty: the patent was granted in 1976 but virtually uncited until around 2019, well after its expiration.
Notably, most of the patent's citations comes from just two companies, Ethicon LLC and Cilag Gmbh International, both of which belong to Johnson \& Johnson.
It appears likely that both the way research is conducted by companies and commercial incentives influence citation behaviors in patents. 
It is all the more interesting to find examples of Sleeping Beauties in patents that are similar to those in law and science.

\section{Citation patterns for APS citation network}
\label{SI:aps-citation-network}

\begin{figure}
    \centering
    \includegraphics[width=\textwidth]{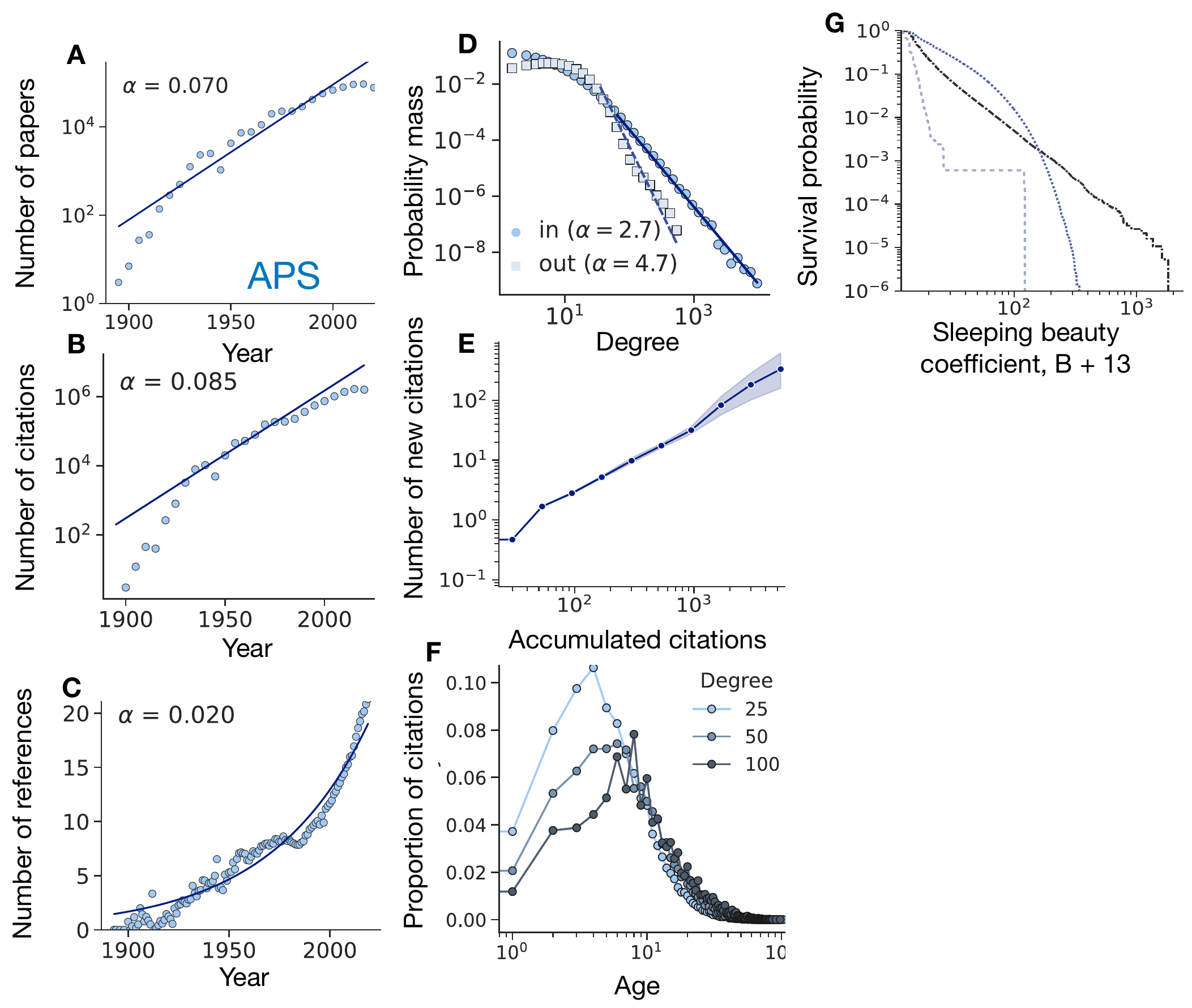}
    \caption{Dynamics of the APS citation network.
    {\bf A}. Growth of publications. 
    {\bf B}. Growth of citations. 
    {\bf C}. Growth of the average number of references. 
    {\bf D}. Degree distribution. 
    {\bf E}. Linear preferential attachment. 
    {\bf F}. Aging function.
    {\bf G}. Skewed distribution of the Sleeping Beauty coefficient. 
    }
    \label{fig:aps-citation-patterns}
\end{figure}

Figure~\ref{fig:aps-citation-patterns} shows the characteristics of citations for the APS dataset. We observe the same qualitative results as for other systems, including preferential attachment, aging, and sleeping beauty publications.

\section{Growth in Citations and References}
\label{SI:ref_growth}

\subsection{The growth of the citations and references}

\begin{figure}
    \centering
    \includegraphics[width=\linewidth]{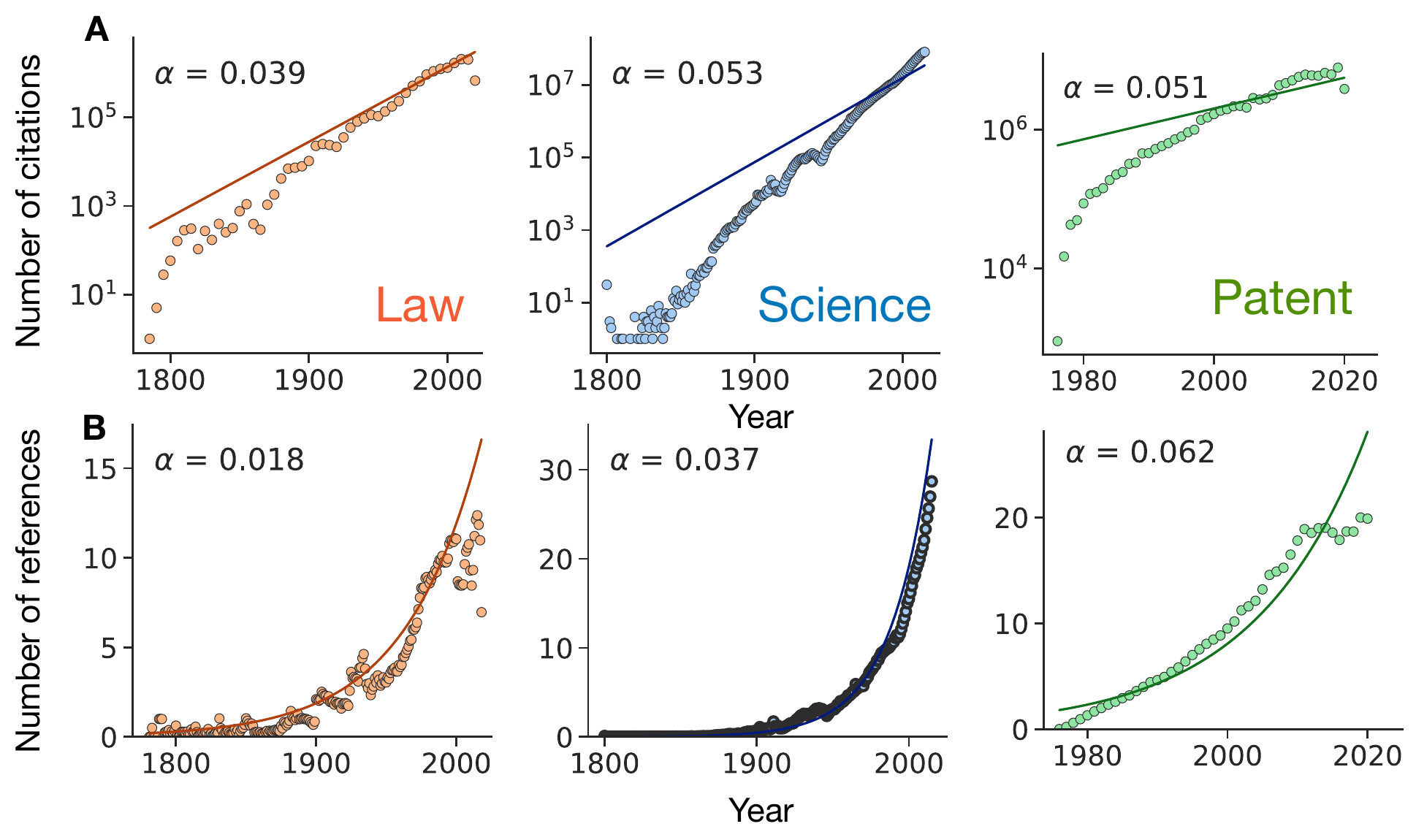}
    \caption{The number of citations and the average number of references per paper over time.}
    \label{fig:growth}
\end{figure}

A common characteristics of the three citations systems  is the growth in citations (Fig.~\ref{fig:growth}). 
We observed that the number of citations, as well as the number of references per paper increases rapidly for the three citation systems. 

\begin{figure}
    \centering
    \includegraphics[width=\linewidth]{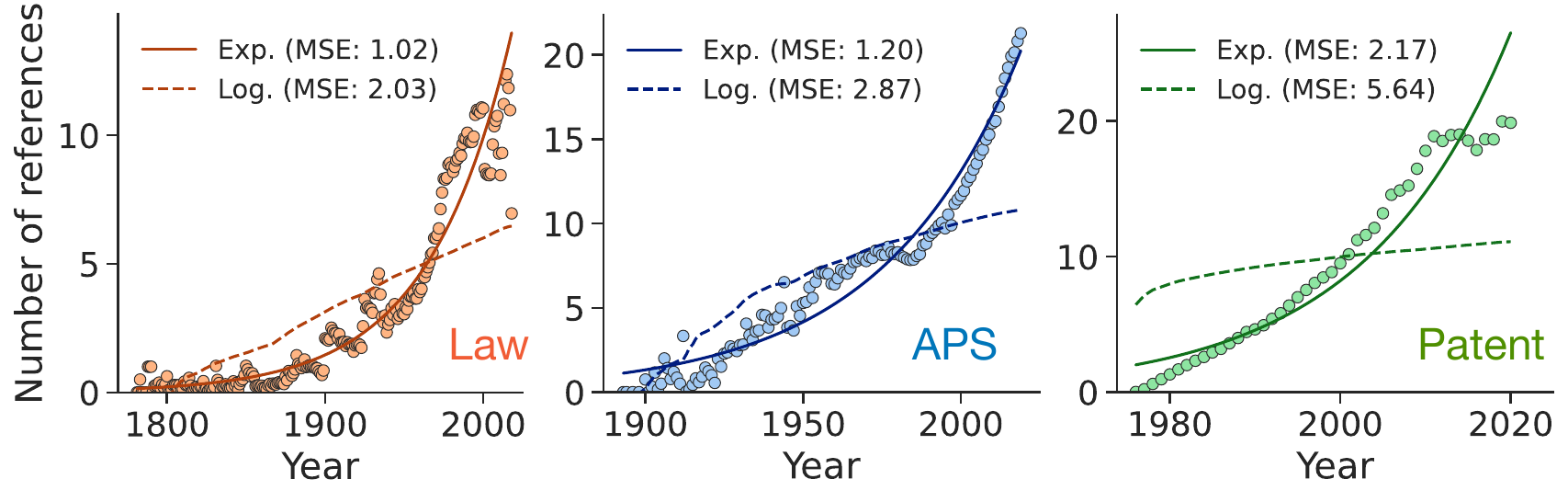}
    \caption{The exponential and logarithmic curve fit for the average number of references per paper. The solid and dashed line represent the exponential and logarithmic curves, respectively.}
    \label{fig:number-of-references-model}
\end{figure}

While the growth in the average number of references appears to be exponential growth, a previous study~\cite{krapivsky2005network} demonstrated that it grows logarithmically.
We observe that the number of references per publication is in fact better characterized by exponential growth.
Specifically, we fit a logarithmic curve of the number $y_r(t)$ of references per publication as a function of the cumulative number $n(t)$ of  publications up to each year, to be in line with the previous study~\cite{krapivsky2005network}, i.e., $y_r(t) = \log(a n(t) + b)$, where $a$ and $b$ are the parameters estimated by minimizing the mean squared error.  
We also fit an exponential curve as a function of the time using the Poisson model (See Section~\ref{sec:exponential_growth}).
Then, we computed the fitting error by the mean squared errors (MSE) for these curves.
We observed that the exponential curve has lower MSE than the logarithmic curve for the APS, law, and patent citation networks (Fig.~\ref{fig:number-of-references-model}). 

\subsection{Citation radius}

\begin{figure}[!ht]
        \includegraphics[width=\hsize]{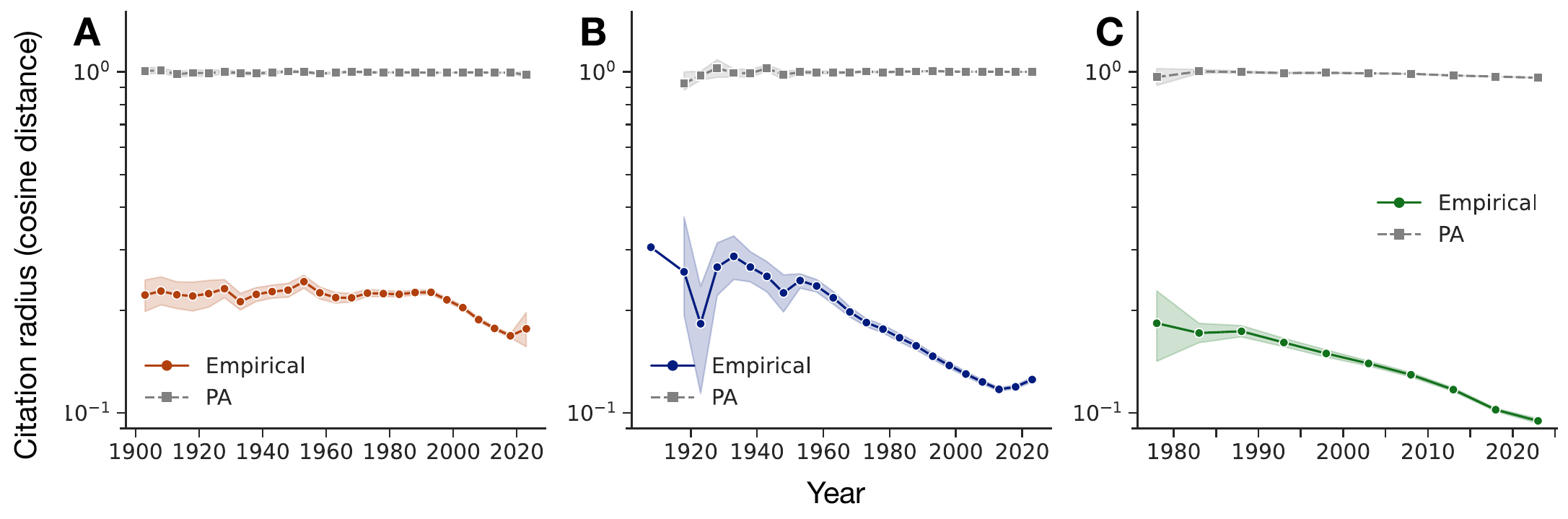}
        \caption{
                The shrinking radius of citation.
                (A) Law.
                (B) APS.
                (C) Patents.
                The radius of references is the average cosine distance from individual publication and its references.
                ``PA'' refers to the reference radius for the CCM fitted on the network generated by the preferential attachment.
            }
        \label{fig:citation_radius}
\end{figure}

By construction, publication-centric baseline models such as the PAM and LTCM cannot account for the exponential growth of references, as they assume a fixed number of $m$ citations per publication.
Within the CCM framework, cited publications (i.e., references) are expected to be thematically adjacent to the citing publication's topic.
In the embedding space, references are therefore, on average, within some small radius of the citing publication's position, where the radius is the average cosine distance between a focal paper and its references.
From this perspective, there are two potential explanations for the increasing number of references over time.
Either the average radius of citation has grown over time, driven, for instance, by increasingly interdisciplinary publications that necessitate broader references.
Alternatively, the average radius of citation has either remained constant or decreased over time, while publications tend to appear in denser areas of the knowledge space, resulting in a growing number of publications available for citation in the vicinity of new publications.
In APS and patent citation networks, we find that the radius of citation---defined by the average distance between a publication and its references---has decreased, suggesting that specialization is increasing (see Fig.~\ref{fig:citation_radius}).
In contrast, we find that the radius of citation in the legal system has remained relatively constant over time.
While increased specialization has been documented in science and patents~\cite{jones2009burden}, it is notable that it does not seem to be as important in law. 
This may be related to the fact that U.S. judges remain generalists. 

\section{Noise contrastive estimation}
\label{SI:model_fitting}

We fit the CCM using the noise contrastive estimation (NCE)~\cite{dyer2014notes,gutmann2010noise}.
NCE is a generic algorithm for fitting an exponential model
\begin{align}
        P(x;\theta) = \frac{\exp(r(x;\theta))}{Z}, \label{eq:NCE_model}
\end{align}
where $x$ is a data sample, and $r(x;\theta)$ is a real-valued function parameterized by $\theta$.
NCE aims to identify the maximum likelihood estimate $\theta^*$ of the model without calculating the computationally-expensive $Z$.
The key idea underlying NCE is to optimize another computationally cheaper function that has the same global maximum at $\theta^*$.
More specifically, NCE fits the parameter $\theta$ by discriminating a data sample $x$ and another ``random'' data $x'$ sampled from an arbitrary distribution $P_0$ using a logistic regression model
\begin{align}
    g(x):=\frac{1}{1 + \exp\left(-r(x;\theta) + \ln P_0(x)  + b\right)},
\end{align}
where $b$ is an intercept that is learned jointly with parameter $\theta$.
NCE trains the logistic regression model by maximizing the log-likelihood
\begin{align}
        \sum_{x \in {\cal X}} \log g(x) - \sum_{x' \in {\cal X}'} \log(1-g(x')), \label{eq:NCE_minima}
\end{align}
where ${\cal X}$ and ${\cal X}'$ are the sets of the given data samples and random data samples generated from $P_0$, respectively.
Notably, Eq.~\eqref{eq:NCE_minima} has a global maximum at MLE $\theta^*$ asymptotically~\cite{gutmann2010noise}.
In other words, the maximization of it provides asymptotically unbiased estimates of Eq.~\eqref{eq:NCE_model}~\cite{gutmann2010noise}.

To train the CCM, we form $x$ by a pair of publications $x=(i,j)$ connected by citations from $i$ to $j$.
Then, we generate a random data point $x'=(i,j')$, where we randomly sample a publication $j'$ published before $i$ with probability $P_0((i,j))\propto c_j(t_i) + c_0$. The unnormalized logarithmic probability for the CCM is given by 
\begin{align}
r((i,j) ; \theta) = \log (c_j(t_i) + c_0) + \log \eta_j - \kappa d(\vec{u}_i, \vec{u}_j)
\end{align}
Then, we train the classifier $g(x)$ given by
\begin{align}
    g(x):=\frac{1}{1 + \exp\left[-\log (c_j(t_i) + c_0) - \log \eta_j + \kappa d(\vec{u}_i, \vec{u}_j)+  \ln P_0((i,j)) + b\right]}.
\end{align}
The intercept $b$ negates the constant terms of $r(i,j)$, i.e., the normalization constant for $P_0((i,j))$, allowing us to simplify it by 
\begin{align}
    g(x):=\frac{1}{1 + \exp\left[- \log \eta_j + \kappa d(\vec{u}_i, \vec{u}_j)+  b\right]}.
\end{align}
To prevent the classifier from overfitting, we regularize it by adding a $L_2$ regularization on $\kappa$.
Thus, we maximize the following likelihood function:
\begin{align}
        \frac{1}{|{\cal X}|}\left(\sum_{(i,j) \in {\cal X}} \log g(i,j) - \sum_{(i,j') \in {\cal X}'} \log(1-g(i,j')) \right) + \frac{\gamma}{2} \kappa^2, \label{eq:NCE_classification}
\end{align}
where $\gamma$ is the regularization parameter, which we set to $5\times10^{-5}$.
We use ADAM~\cite{kingma2014adam} with batch size of $4096$ edges, learning rate $10^{-3}$, and $120$ epochs.

We generate a citation network using the parameters estimated based on the empirical citation network.
The generation process is akin to the preferential attachment, with the probability of citations given by Eq.~\eqref{eq:CCM}.
In each year $t$, we add $n_t$ new publications, where $n_t$ is the number of publications that appear in year $t$ in the empirical network.
Each new publication $i$ has the same number $m_i$ of references as that in the empirical network.
Each edge is placed with probability proportional to Eq.~\eqref{eq:CCM}.
The resulting network has the same number of nodes, edges, and growth rate as the empirical network.

\section{Number of parameters of citation models}
\label{sec:model_size}

In the PAM, LTCM, and CCM, the likelihood of receiving citations is proportional to the total number of accumulated citations plus an additional offset citation $c_0$, which ensures that newly published publications without any citations have a chance to receive citations.
Additionally, the LTCM has three parameters per publication: fitness ($\eta_j$) and aging ($\mu_j$, $\sigma_j$). 
This results in $3N$ parameters.
CCM has $K$ parameters per publication: $\eta_j$ and the $K$-dimensional embedding vector $\vec{u}_j$. 
Note that, since the embedding vector is unit length, the effective number of parameters for an embedding of a publication is $K-1$. 
CCM also has scaling parameters ($\kappa$ and $\lambda$). 
Consequently, CCM has $KN + 2$ parameters in total.

\section{Robustness of the results}
\label{sec:sensitivity_analysis_reproduceability}

We validated the robustness of the CCM by changing the model's hyperparameters $(c_0, K)$.
Overall, the CCM is robust to the choice of hyperparameters, except for the number of dimensions $K$ of the embedding space, which must be larger then a minimum threshold (Figs.~\ref{fig:sensitivity_c0} and \ref{fig:sensitivity_dimension}).
We also explore the effect of aging and fitness by comparing the CCM with and without aging and fitness.
We identify that the current configuration of the CCM is minimal yet sufficient to reproduce the citation dynamics observed in the three knowledge systems (Fig.~\ref{fig:sensitivity_aging_factor}, \ref{fig:prediction_aging_factor}, \ref{fig:sensitivity_fitness}, and \ref{fig:prediction_fitness}).
We discuss these results in detail in the following sections, in each of which we will first show the reproducibility of the citation dynamics and then show predictability of the citation dynamics.
We used the APS dataset to examine various configurations of the CCM, except for the validation in Section~\ref{sec:sensitivity_analysis_prediction_percentile}, where we assessed performance at different citation percentiles because this validation was feasible for the SciSciNet dataset.

\subsection{Embedding dimension and offset citations}
\label{sec:sensitivity_analysis_hyperparameters}

\subsubsection{Citation pattern reproducibility}

\begin{figure}
    \centering
    \includegraphics[width=\hsize]{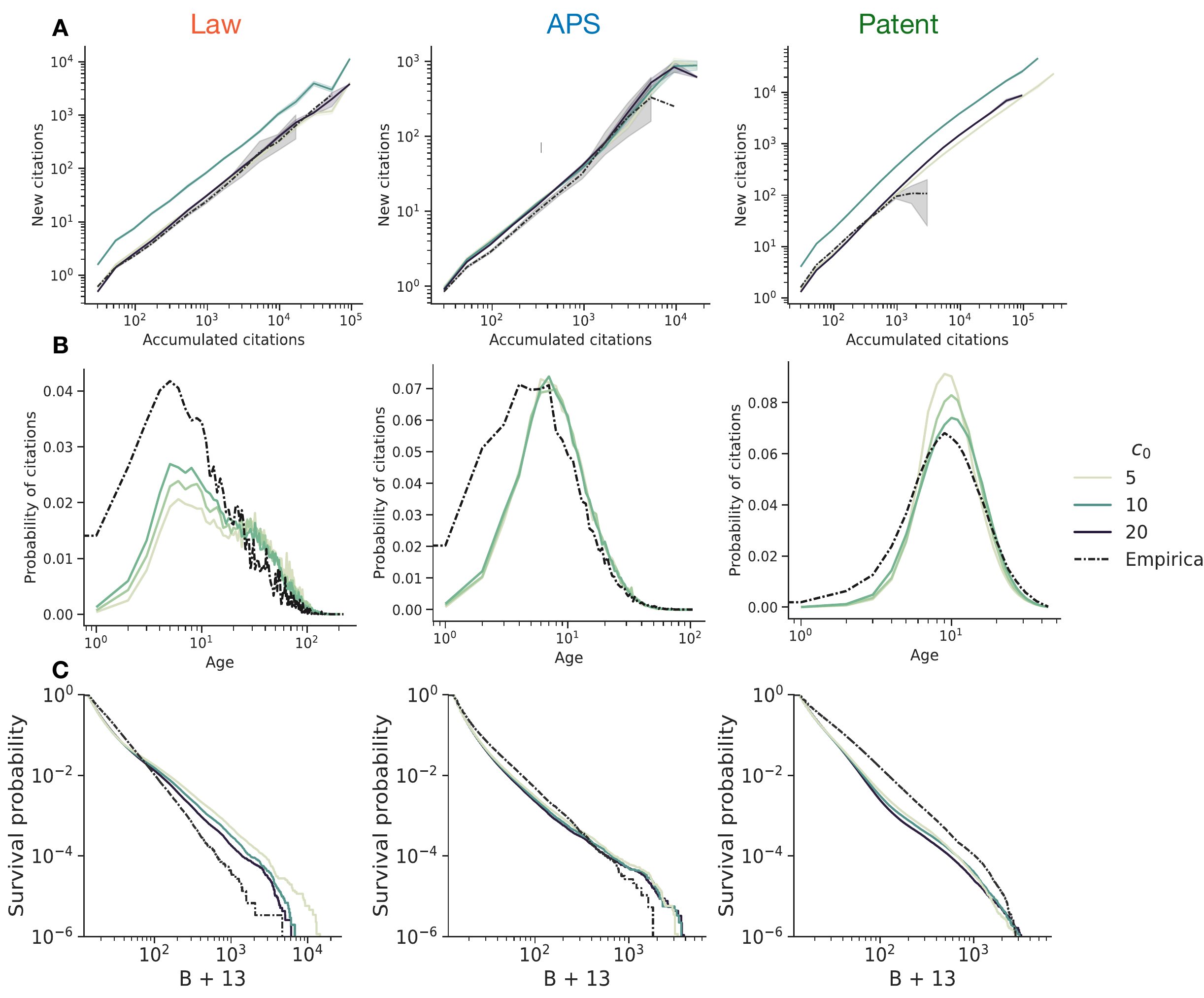}
    \caption{Citation patterns for the networks generated by the CCM with different values of $c_0$.}
    \label{fig:sensitivity_c0}
\end{figure}

\begin{figure}
    \centering
    \includegraphics[width=\hsize]{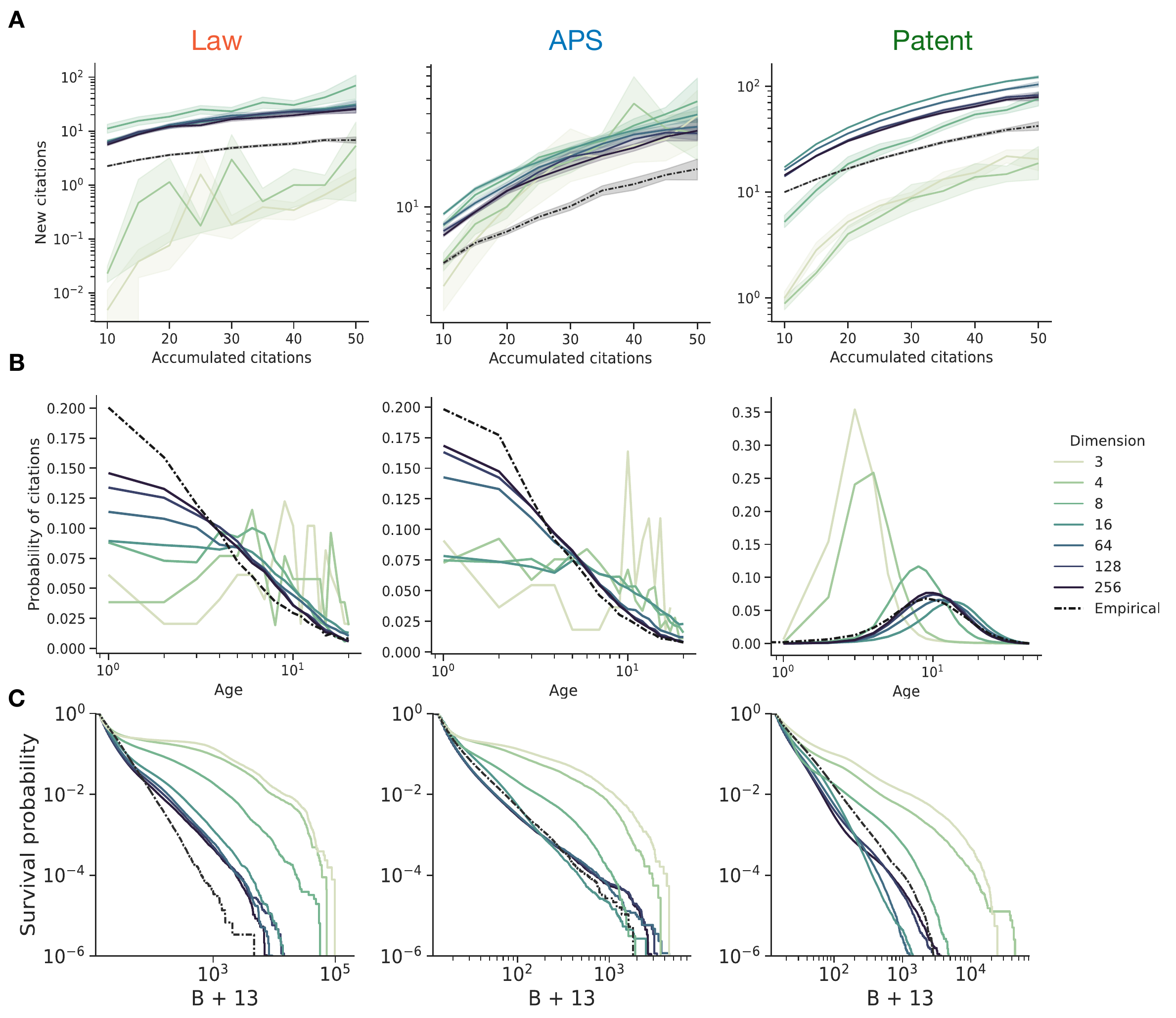}
    \caption{Citation patterns for the networks generated by the CCM with different values of $K$.}
    \label{fig:sensitivity_dimension}
\end{figure}

CCM has two hyperparameters, namely the offset citations $c_0$ and the number $K$ of dimension of the embedding.
We test if three citation patterns---preferential attachment, recency, and heterogeneous SB distribution---are reproduced for different values of $c_0$ and $K$.
First, the CCM reproduces the three citation patterns qualitatively similarly to the empirical ones for different values of $c_0$ (Fig.~\ref{fig:sensitivity_c0}).
Second, while the number $K$ of dimensions does not strongly affect preferential attachment, different dimensions can produce substantially different recency biases and SB distributions (Fig.~\ref{fig:sensitivity_dimension}).
Specifically, the recency bias is reproduced only if the embedding space is sufficiently high dimensional, likely because communities cannot drift away from the publications in a low-dimensional embedding space. 
For low-dimensional embedding spaces ($K \leq 16$), the sleeping beauty distributions have a substantial margin to the empirical network, while the margin is closed for higher dimensional embedding spaces ($K \geq 64$).
In summary, the CCM is robust to the choice of $c_0$ and $K$, provided that $K$ is larger than a minimum threshold.

\subsubsection{Citation predictability}

The CCM is a rich citation model compared to other existing models.
Thus, it is possible that the high predictability results from the richness of the citation model. 
We test a smaller CCM with the same order of parameters as the LTCM and compare their prediction performances.

\begin{figure}
    \centering
    \includegraphics[width=\hsize]{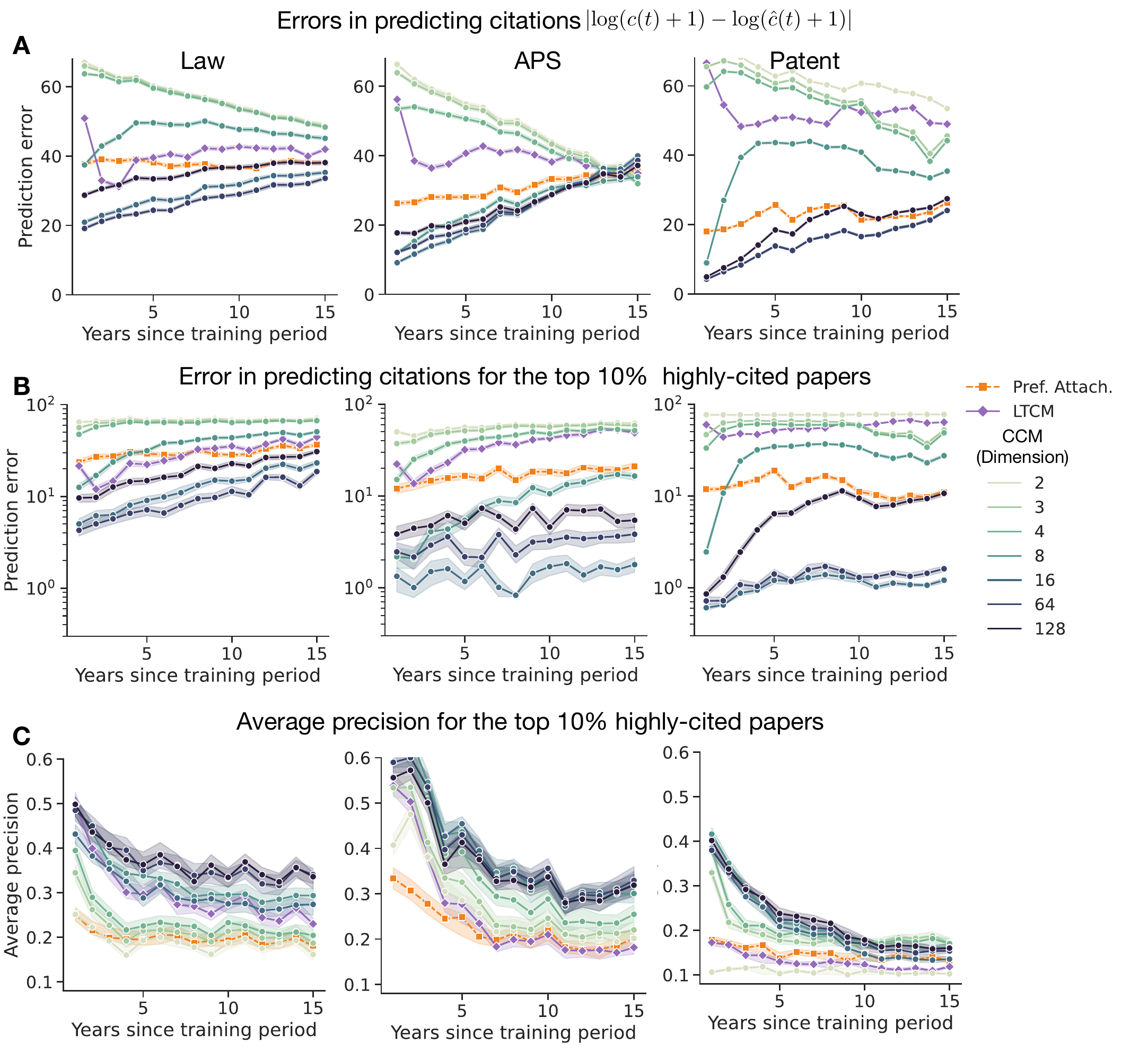}
    \caption{Prediction of citations for different embeddings dimensions $K$.}
    \label{fig:prediction_dim}
\end{figure}

Figure~\ref{fig:prediction_dim} shows the prediction performance of the CCM model for different values of $K$.
Overall, a higher dimension $K$ yields a better prediction performance, in terms of predicting both the citation count and highly-cited publications.
We note that a smaller $K$ sometimes yields a better prediction performance; for example, the CCM with $K=16$ dimensional space yields the lowest prediction error (Figs.~\ref{fig:prediction_dim}A and B).
This result implies that, when $K$ is excessively large, CCM may suffer from overfitting to the training data, thereby undermining its ability to accurately predict unseen test data.
Therefore, it is not always the case that a richer model always yields better predictions.

We compare the CCM with $K=3$ to the LTCM, which has approximately the same number of parameters.
The LTCM predicts the citations better than the small CCM  for the first ten years from the publications but worse in the long-term (10$<$ years) for science and patent citations (Figs.~\ref{fig:prediction_dim}A and B).
While the CCM with $K=3$ does not accurately predict the citation counts, it does a good job at predicting \emph{which} publications will be highly cited (Fig.~\ref{fig:prediction_dim}C).

\subsection{Aging}
\label{sec:sensitivity_analysis_aging}

\subsubsection{Citation pattern reproducibility}

The CCM does not include the aging factor (i.e., $f(t_i - t_j\;;\mu_j, \sigma_j)$), which we identified to be an unnecessary component.
Furthermore, we find that including the aging factor in the CCM results in a noticeable decrease in the number of sleeping beauties.

\begin{figure}
    \centering
    \includegraphics[width=\hsize]{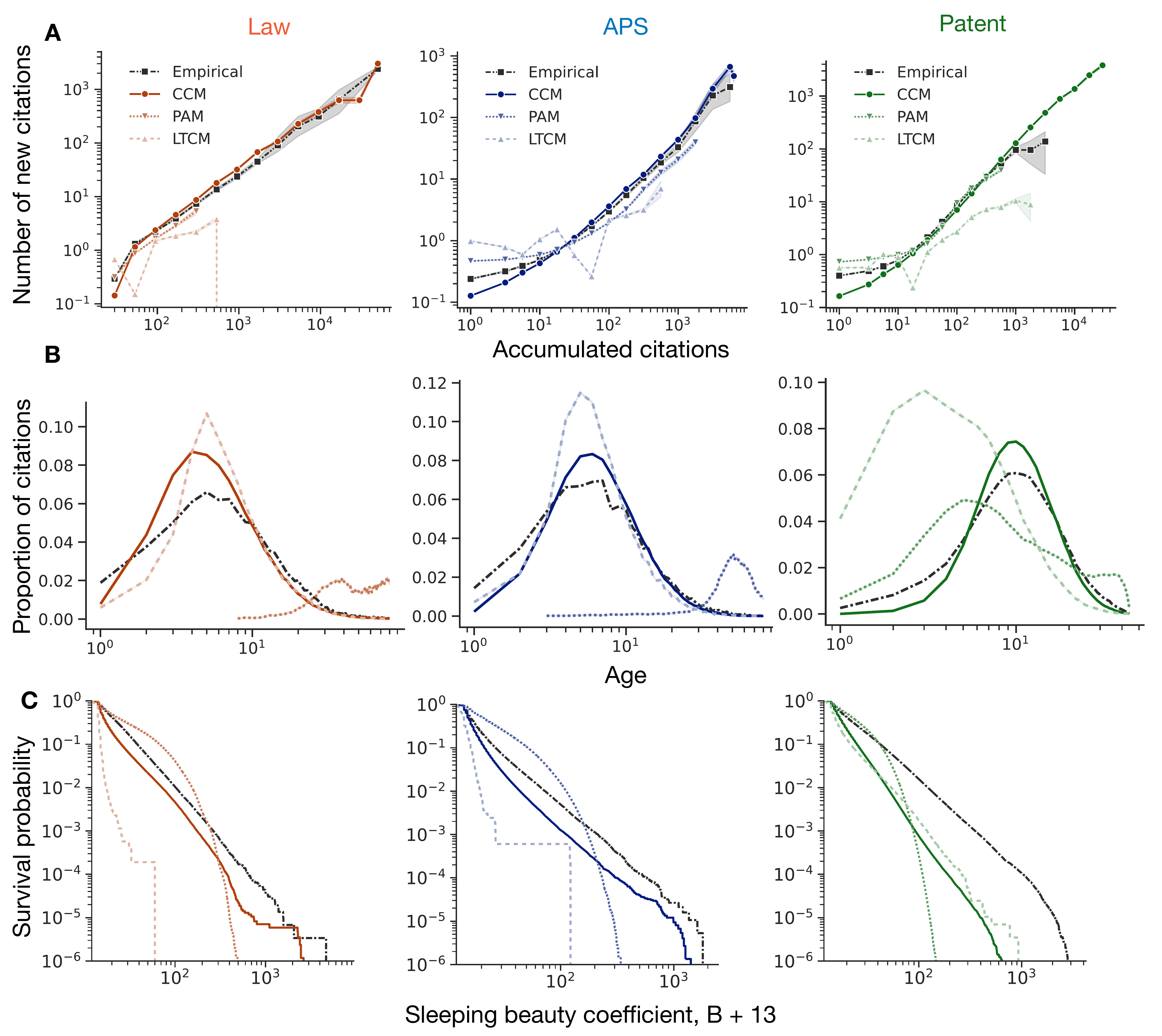}
    \caption{Citation patterns for the networks generated by the CCM with the aging factor given by Eq.~\eqref{eq:CCM_aging}.}
    \label{fig:sensitivity_aging_factor}
\end{figure}

We define the CCM with aging as follows:
\begin{align}
    \label{eq:CCM_aging}
    P_{\text{CCMa}}(j \given i; \vec{u}_i, \vec{u}_j) = \frac{1}{Z(t_j)} \left(c_{j} + c_0\right) \eta_j f(t_i - t_j\;;\mu, \sigma) \exp\left(-\kappa d(\vec{u}_i,\vec{u}_j)\right),
\end{align}
where we use the same aging function $f$ (i.e., the log-normal distribution) as for the LTCM with parameters $\mu$ and $\sigma$ being equal across all publications.
The CCM with the aging factor qualitatively reproduces preferential attachment and recency patterns, but it produces fewer sleeping beauties, especially in the patent citation network (Fig.~\ref{fig:sensitivity_aging_factor}).

\subsubsection{Citation predictability}

\begin{figure}
    \centering
    \includegraphics[width=\hsize]{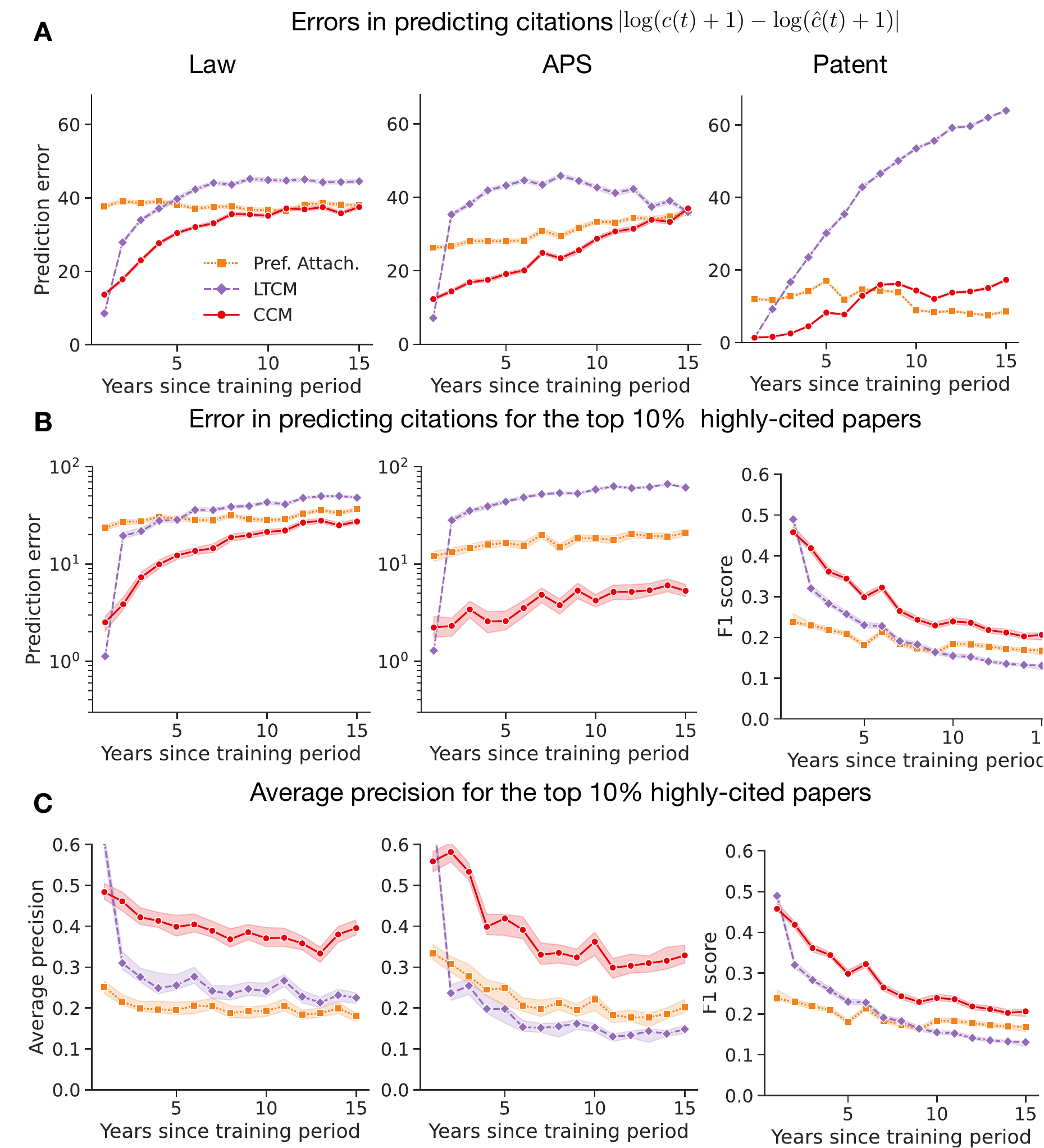}
    \caption{Prediction of citations for the publications published in 2000 by the CCM with the aging factor given by Eq.~\eqref{eq:CCM_aging}.}
    \label{fig:prediction_aging_factor}
\end{figure}

We find that the inclusion of the aging factor can slightly enhance the citation predictability of the CCM, especially the citations of young publications (Fig.~\ref{fig:prediction_aging_factor}).
This suggests that considering the temporal dynamics of citation can improve the model's ability to predict future citations.

\subsection{Fitness}
\label{sec:sensitivity_analysis_fitness}

\subsubsection{Citation pattern reproducibility}

\begin{figure}
    \centering
    \includegraphics[width=\hsize]{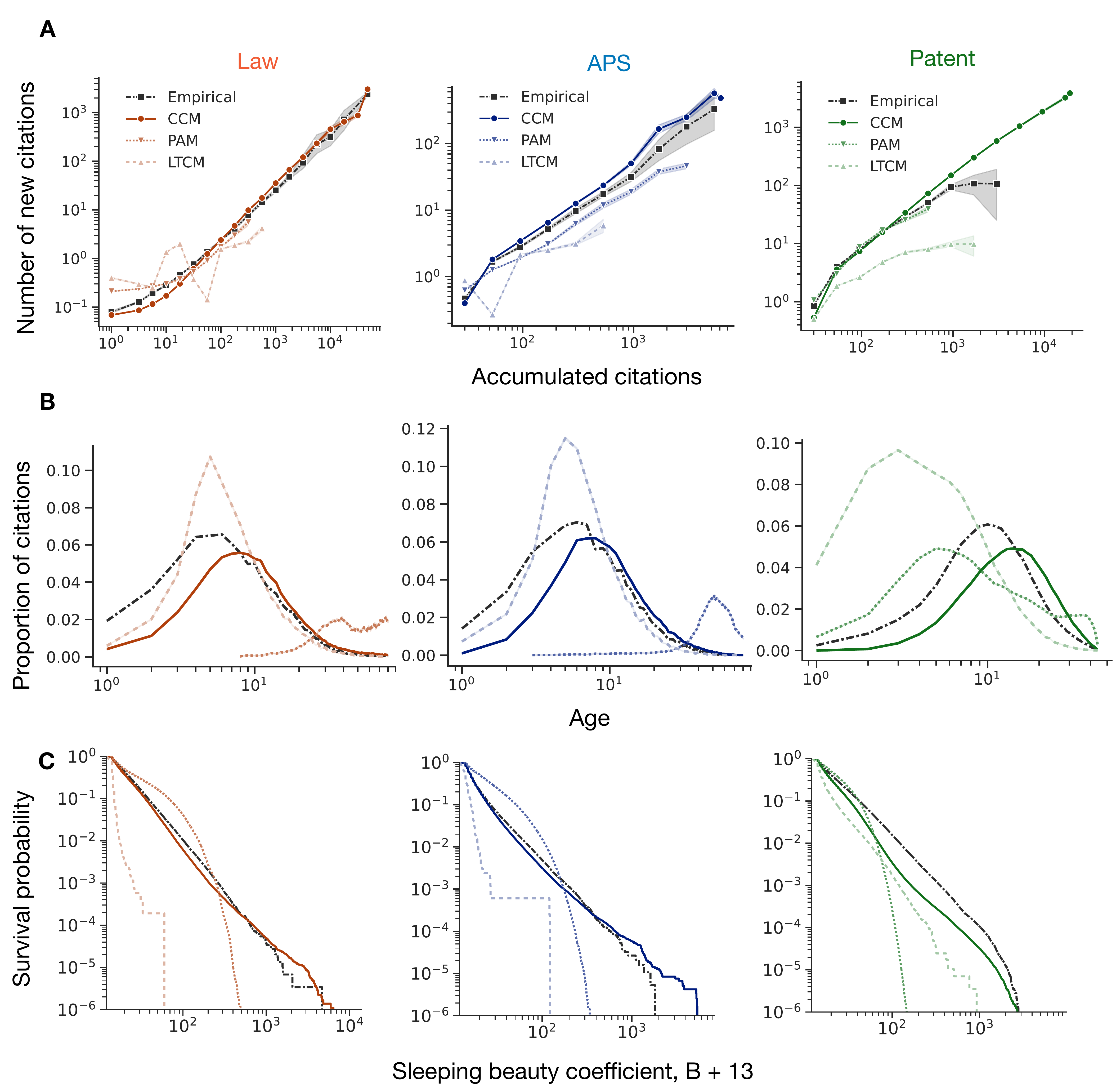}
    \caption{Citation patterns for the networks by the CCM without the fitness factor given by Eq.~\eqref{eq:CCM_wo_fitness}.}
    \label{fig:sensitivity_fitness}
\end{figure}

We define the CCM without the fitness factor as follows:
\begin{align}
    \label{eq:CCM_wo_fitness}
    P_{\text{CCMw}}(j \given i; \vec{u}_i, \vec{u}_j) = \frac{1}{Z(t_j)} \left(c_{j} + c_0\right) \exp\left(-\kappa d(\vec{u}_i, \vec{u}_j)\right),
\end{align}
where we omit the fitness parameter $\eta_j$ from the original CCM (Eq.~\eqref{eq:CCM}).
The CCM without the fitness factor qualitatively reproduced all the citation characteristics of the empirical networks, including preferential attachment, recency, and a heterogeneous SB coefficient distribution (Fig.~\ref{fig:sensitivity_fitness}).

\subsubsection{Citation predictability}

\begin{figure}
    \centering
    \includegraphics[width=\hsize]{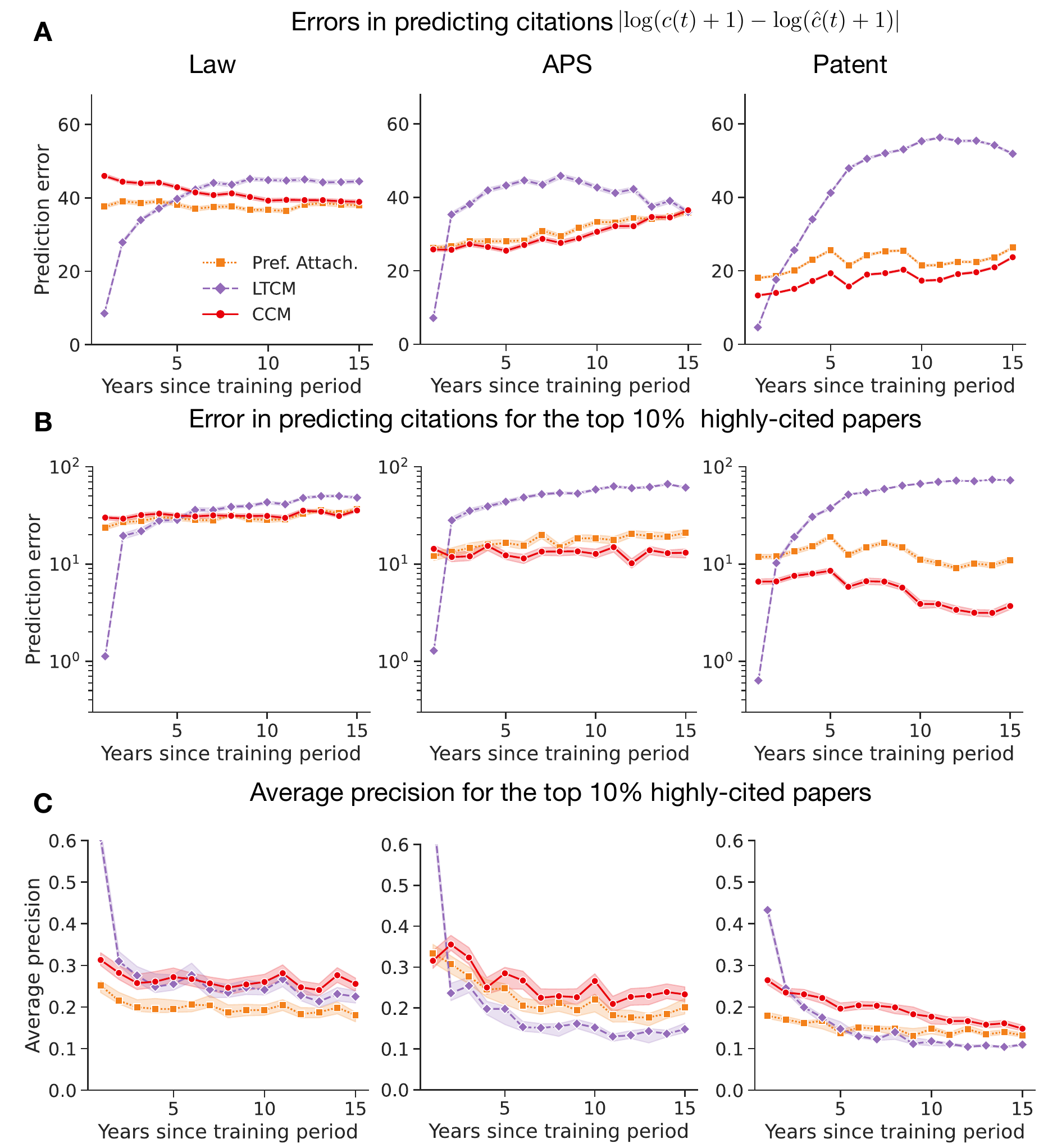}
    \caption{Prediction of citations for publications published in 2000 by the CCM without fitness factor given by Eq.~\eqref{eq:CCM_wo_fitness}.}
    \label{fig:prediction_fitness}
\end{figure}

When it comes to predicting citations, the CCM without a fitness factor significantly underperforms the CCM with the fitness factor, and performs worse than the LTCM for the legal citation networks (Fig.~\ref{fig:prediction_fitness}).

\subsection{Different percentile of the highly-cited papers}
\label{sec:sensitivity_analysis_prediction_percentile}

\begin{figure}
    \centering
    \includegraphics[width=\hsize]{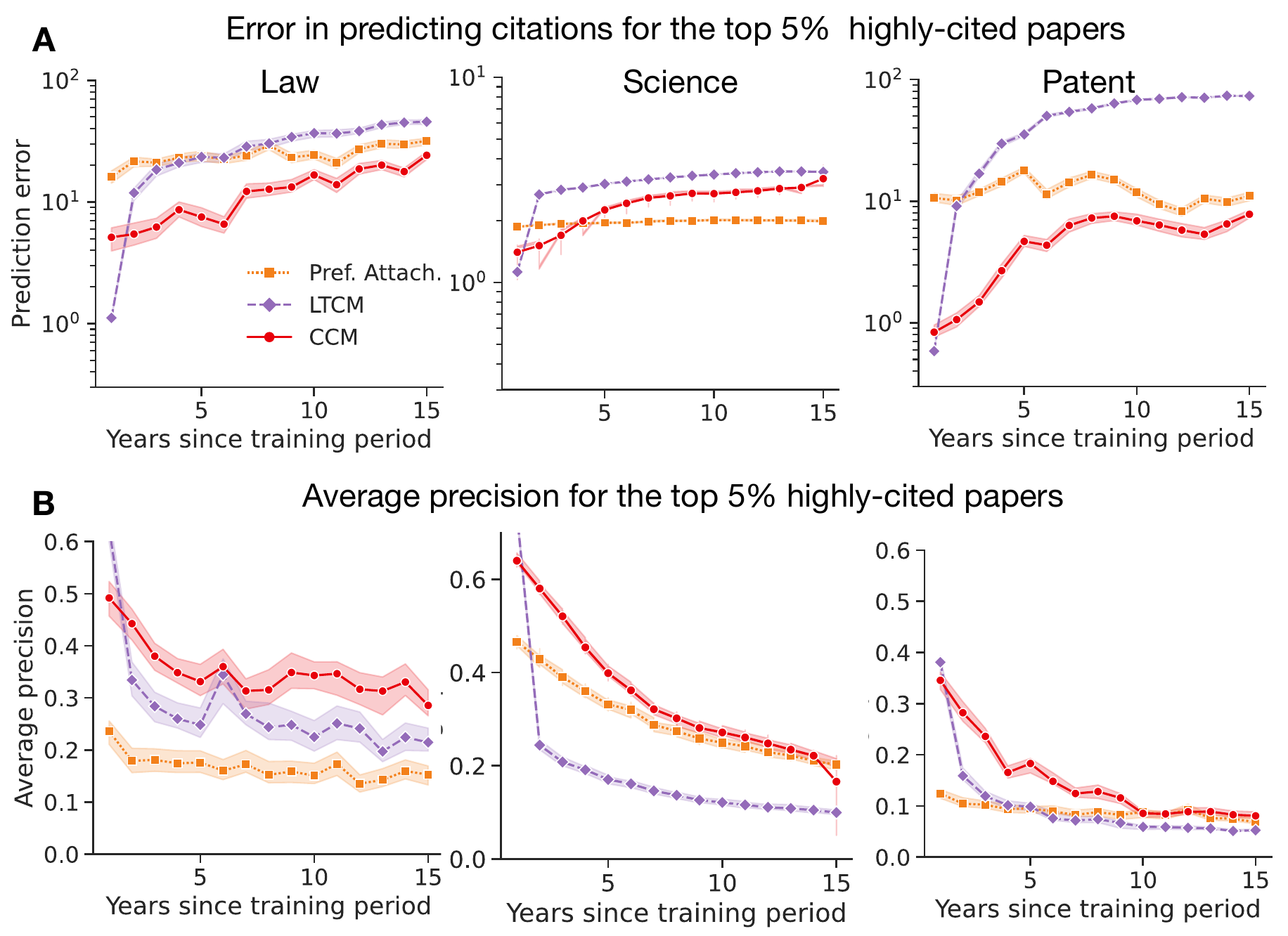}
    \caption{Predictions for the top-5\% highly cited publications.}
    \label{fig:top_5_prediction}
\end{figure}

We test the citation models by focusing on the top-$5\%$ most highly cited publications (Fig.~\ref{fig:top_5_prediction}).
As for the top-10\% highly-cited papers, the CCM outperforms the LTCM for the top-5\% highly-cited papers.

\subsection{Different cohorts of publications for predictions}
\label{SI:sensitivity_analysis_prediction_cohort}

\begin{figure}
    \centering
    \includegraphics[width=\hsize]{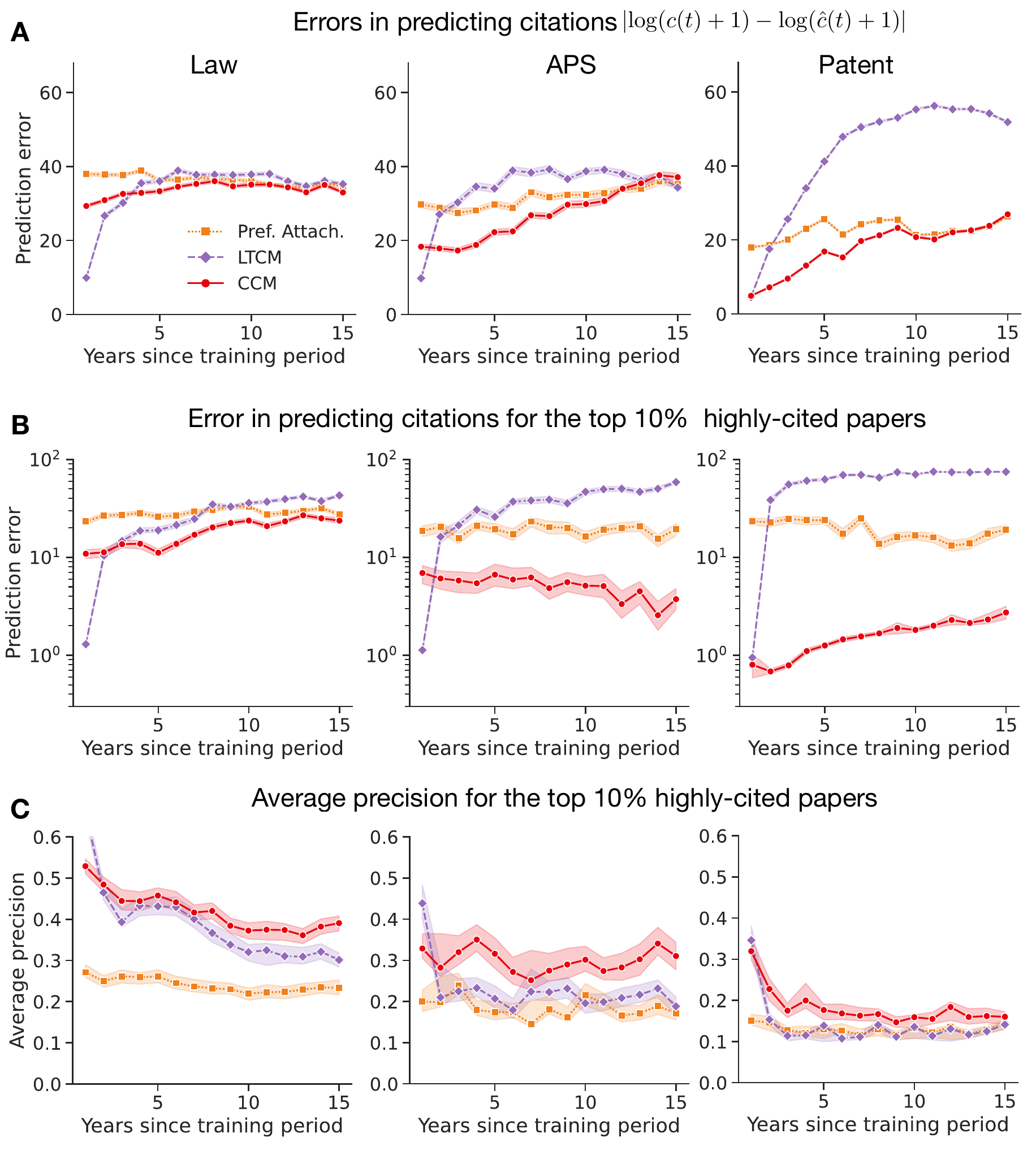}
    \caption{Prediction for the publications published in 1990.}
    \label{fig:prediction_1990}
\end{figure}

We test the citation models by using different cohorts of publications for predictions.
Specifically, we focus on publication published in 1990, with five years of citation data for the models to learn from.
We find qualitatively similar results as reported for publications published in 2000 (Fig.~\ref{fig:prediction_1990}).

\end{document}